\documentstyle[aps]{revtex}
\input{epsf}

\begin{document}

\title{Theory of Vibrationally Inelastic Electron Transport\\ 
       through Molecular Bridges}

\author{M.\ {\v{C}\'{\i}\v{z}ek}\footnote{%
            Permanent address: Institute for Theoretical Physics,
            V Hole\v{s}ovi\v{c}k\'{a}ch 2, 180 00 Praha, Czech
            Republic},
        M.\ Thoss, and W.\ Domcke}
\address{
  Institute of Physical and Theoretical Chemistry,\\
  Technical University of Munich,\\
  D-85747 Garching, Germany\\
}
\date{\today}
\maketitle

\vspace*{-0.4cm}

\begin{abstract}

\vspace*{-0.4cm}

Vibrationally inelastic electron transport through a molecular bridge that is
connected to two leads is investigated. 
The study is based on a generic model of vibrational excitation 
in resonant transmission of electrons  through a molecular junction.
Employing methods from electron-molecule scattering theory,
the transmittance through the molecular bridge can be evaluated numerically exactly.
The current through the junction is obtained approximately using a Landauer-type formula.
Considering different parameter regimes, which include both the case of a molecular bridge that
is weakly coupled to the leads, resulting in narrow resonance structures, and the 
opposite case of a broad resonance caused by strong interaction with the leads,
we investigate the characteristic effects of coherent and dissipative vibrational motion 
on the electron transport. Furthermore, the validity of widely used approximations 
such as the wide-band approximation and the restriction to elastic transport mechanisms 
is investigated in some detail.
\end{abstract}


\section{Introduction}
\label{sec:introduction}

The experimental demonstration of the possibility 
to connect two electrodes by a single molecule 
and to measure a current through such a molecular junction
\cite{rzmbt97,ruitenbeck02,reichert02} has stimulated increasing theoretical efforts
to elucidate the basic mechanisms of electron transport in such systems 
(see, for example, \cite{joachim00,n01,grossmann02,nr03,hry02} and references therein). 
Most of the theoretical work in recent years has 
been devoted to the determination of the electronic structure of 
molecular junctions, employing a variety of  methods that include
extended H\"uckel approaches \cite{dthrhk97,ek98a,tdhrhk98,mj97,mj97b,mj97c}, 
{\em ab initio} quantum-chemistry methods \cite{fclg97,lk99,yrgmr99}, 
and density functional theory \cite{l95,ht95,la00,vpl00,mrr00,lb03}. 
The majority of these studies have focused
on elastic mechanisms for electron transport, where the current through
the molecular junction can be obtained from the single-electron transmission probability
using the Landauer formula \cite{landauer57,landauer81,datta}. These studies
have demonstrated the importance of the electronic energy level structure
of the molecular bridge: the tunneling of electrons through occupied and unoccupied
levels of the molecule results in resonance structures in the transmission
probability which in turn may cause strongly nonlinear current-voltage characteristics.
 
Much less is known about the effect of vibrationally inelastic processes,
associated with the vibrational motion of the molecular bridge, on the electron transport.
In experiments on electron transport through ${\rm H}_2$ molecules between two platinum 
electrodes \cite{ruitenbeck02} as well 
as  C$_{60}$ molecules connected to gold electrodes \cite{pplaam00}, 
indications for an influence of the
center-of-mass motion of the respective molecule on the conductivity have been found. 
Effects of the internal vibrational motion of the molecular bridge on the current 
through the junction, on the other hand,  have (to our knowledge) not yet  been reported. 
Such effects have, however, been predicted in a variety of theoretical studies.
For example, the "static"  influence of  the internal vibrational modes  has been studied by averaging the
transmittance over the probability distribution of the vibrational degrees of freedom \cite{omwkrlm98,s03,trn03}.
The dynamical impact of the vibrational degrees of freedom on the  
tunneling current in molecular junctions has been investigated within, 
e.g., nearest neighbor tight-binding models \cite{ek00,w03,bs01,mpt03}.
The effect of the center-of-mass motion on the current has been explored 
for transversal vibrations \cite{bf04} and for longitudinal vibrations 
(shuttling mechanism) \cite{givksj98,ndj03}.
These studies have demonstrated that the vibrational
motion of the molecular bridge may result in additional (vibrational) resonance structures in 
the transmission probability which can alter the current-voltage characteristic significantly.
Furthermore, the excitation of the vibrational degrees of freedom of the molecule
provides a mechanism for heating of the molecular junction and thus is a possible source 
of instability \cite{segal02,segal03}.

The effect of vibrationally inelastic processes on electron transport has also 
been investigated for a variety of closely related problems including
the tunneling of electrons through long 
polymer chains (molecular wires) \cite{yssb99,nf99}, 
electron transport through quantum dots and heterostructures \cite{d84,shls91,hb99,wjw88,wjw89},
as well as the theoretical description of single-molecule vibrational spectroscopy 
in scanning tunneling microscopy (STM) experiments
\cite{pb87,lp00,sb97,ga93,bhg93,bh94,gpl97,ga99}. 
The formally related process of electron transport in the presence of a
laser field has also been studied \cite{clkh03,lkhn03}.

Another closely related process is vibrationally inelastic electron-molecule scattering.
Here, it is well established that the resonant scattering of a low-energy electron 
from a molecule can result in strong vibrational excitation. This process has been
studied in great detail experimentally (see, for example, the reviews 
\cite{s73b,a89}). Furthermore, efficient theoretical methods have been developed 
to describe the interaction of electronic
and vibrational degrees of freedom in resonant collision processes of low-energy electrons with molecules 
\cite{d91}.
As a result, inelastic electron scattering from diatomic molecules
is now well understood from first principles \cite{chad02}.
Due to the close similarities between the process of vibrational excitation induced
by electron scattering from a molecule
and vibrationally inelastic electron transmission trough a molecular junction, 
it is to be expected that the 
methods and concepts developed in the former field can advantageously be 
used in the latter field.
An example is the so-called wide-band approximation, where the
energy dependence of the coupling between the molecule and the leads is
neglected. This approximation, which has been adopted in most of the theoretical studies
of vibrationally inelastic electron transport in molecular junctions
(for exceptions, see  
Refs.\ \cite{ek00,nf99,shls91,hb99,gsl89})
has been tested in detail for resonant electron-molecule scattering \cite{d91}.
It has been shown that the wide-band approximation breaks down at energies close
to thresholds and often does not accurately describe vibrational excitation processes.

In this paper we study vibrationally inelastic effects on electron
transport through a molecular junction beyond the wide-band approximation. To this end, we consider a generic model for vibrational excitation
in resonant electron transmission processes trough a molecular junction.
The model includes the coupling of an electronic resonance state, 
located at the molecular bridge,
to the continuum of electronic lead
states as well as the coupling of the electronic degrees of freedom to a 
vibrational reaction mode of the molecule.
Employing projection-operator techniques \cite{Feshbach62}
well-known from electron-molecule scattering \cite{d91},
the transmission probability through the molecular junction can be evaluated
numerically exactly within this model. 
The current trough the bridge is obtained
employing a generalized Landauer formula \cite{note1}.
Based on numerical results for models in different parameter
regimes, we  
study the importance of inelastic effects on molecular conductance
as well as the
validity of the wide-band approximation.

Furthermore, we investigate how vibrationally inelastic  effects on electron transport
are altered if the vibrational motion has dissipative character. 
Dissipative vibrational processes
(such as, for example, vibrational dephasing and relaxation) are 
expected to be of importance in larger molecules or in molecular bridges that are embedded
in an environment.
To describe dissipative vibrational motion, we consider the coupling of the reaction
mode to a vibrational bath. The various observables are then obtained employing an 
expansion with respect
to the number of quanta in the final state of the bath. This technique, 
which has been proposed some years ago 
in the context of electron scattering from large molecules 
\cite{td98}, 
allows us 
to describe the effect of vibrational relaxation in an approximate, yet controlled, way,
without invoking  Markov-type approximations.
Moreover, it is shown that in the case of  identical left and right leads
and zero bias voltage
a unitarity condition can be exploited, which allows a numerically
exact evaluation of the transmission probability, including the effects of a 
dissipative vibrational bath.

This paper is organized as follows: 
After an introduction of the model and the observables of interest, Sec.\ \ref{sec:model}
outlines the theoretical methods used to describe the transmission probability and the current 
through the molecular  bridge. In particular, we discuss various levels of the theoretical treatment:
a theoretical description based on purely elastic transport mechanisms, 
the incorporation of vibrationally inelastic processes (coherent and dissipative),
as well as the wide-band approximation.
Sec.\ \ref{sec:results} presents model studies of vibrationally inelastic electron
transport for different parameter regimes, comprising 
both the case of a molecular bridge that
is weakly coupled to the leads, resulting in narrow resonance structures, and the 
opposite case of a broad resonance caused by strong interaction with the leads.
Furthermore, 
the various levels of theory are critically compared.
Finally, Sec.\ \ref{sec:conclusion} gives a summary and concludes.


\section{Theory}
\label{sec:model}

\subsection{Model Hamiltonian}

To investigate the influence of vibrational motion on the transmission of 
electrons through a molecular bridge,
we consider a situation where 
two metallic leads, which serve as a reservoir of electrons,
are connected to a molecule through which electrons can be transferred 
from one lead the other.
As has been demonstrated in previous work on elastic electron transport 
(see, for example, \cite{kbpevmj99}), 
the transmission of electrons through molecular junctions is typically characterized by resonances
which correspond to the various electronic orbitals of the bridging molecule.
From the theoretical point of view, the situation is thus 
characterized by a set of resonance states 
which are embedded in the continuum of lead states.
In this paper, we will consider, for simplicity, a situation where
only a single electronic resonance, corresponding (in the limit of vanishing coupling to the leads)
to a molecular anion, contributes to the transmission process.

From the theory of resonant electron-molecule scattering, it is well
known that the influence of vibrational motion on the electron transmission 
can be advantageously described by choosing a basis
of diabatic electronic states  consisting of  a discrete state $|\phi_{\rm d}\rangle$,
which represents the resonance (i.e.\ the situation where the transmitting
electron is situated  at the bridging molecule) 
and a set of orthogonal continuum states, $|\phi_{k\alpha}\rangle$, 
$\alpha=$L,R, describing the electron in the left
and right lead, respectively. 
Accordingly, the Hamiltonian reads
\begin{equation}\label{e:defhs}
  H_{\rm S}=|\phi_{\rm d}\rangle \tilde H_{\rm d} \langle\phi_{\rm d}|
   + \sum_{k,\alpha=L,R}
    \left\{
       |\phi_{k\alpha}\rangle (\epsilon_{k\alpha}+\tilde H_0)\langle\phi_{k\alpha}|
      +|\phi_{\rm d}\rangle V_{{\rm d}k\alpha}\langle\phi_{k\alpha}|
      +|\phi_{k\alpha}\rangle V_{{\rm d}k\alpha}^*\langle\phi_{\rm d}|
    \right\},
\end{equation}
where $\tilde H_0$ denotes the vibrational Hamiltonian of the neutral molecule
in the electronic ground state and $\tilde H_{\rm d}$ the vibrational
Hamiltonian in the discrete electronic state $|\phi_{\rm d}\rangle$. 
The electronic coupling between the leads and the molecule is specified by the
coupling matrix elements  $V_{{\rm{d}}k\alpha}$.

\begin{figure} 
 \begin{center} 
  \setlength{\unitlength}{3ex} 
  \begin{picture}(29.0,6.0)(1.0,1.5) 
   \multiput(3.5,3.5)(4.0,0.0){3}{\circle{1.0}} 
   \put(15.5,3.5){\circle{1.5}} 
   \multiput(19.5,3.5)(4.0,0.0){3}{\circle{1.0}} 
   \thicklines 
   \multiput(4.0,3.5)(4.0,0.0){2}{\line(1,0){3.0}} 
   \multiput(20.0,3.5)(4.0,0.0){2}{\line(1,0){3.0}} 
   \multiput(1.5,3.5)(26.5,0.0){2}{\line(1,0){1.5}} 
   \thinlines 
   \multiput(12.0,3.5)(4.25,0.0){2}{\line(1,0){2.75}} 
   \multiput(3.3,2.0)(4.0,0.0){3}{$\mu_{\rm L}$} 
   \multiput(19.3,2.0)(4.0,0.0){3}{$\mu_{\rm R}$} 
   \put(15.3,2.0){$\epsilon_{\rm d}$} 
   \multiput(5.2,4.0)(4.0,0.0){2}{$\beta$} 
   \multiput(21.2,4.0)(4.0,0.0){2}{$\beta$} 
   \put(13.2,4.0){$v$} 
   \put(17.2,4.0){$v$} 
   \put(3.2,3.35){-$\scriptstyle 3$} 
   \put(7.2,3.35){-$\scriptstyle 2$} 
   \put(11.2,3.35){-$\scriptstyle 1$} 
   \put(15.1,3.35){$\scriptstyle|\phi_d\rangle$} 
   \put(19.4,3.35){$\scriptstyle 1$} 
   \put(23.4,3.35){$\scriptstyle 2$} 
   \put(27.4,3.35){$\scriptstyle 3$} 
   \put(4.0,6.0){Left lead} 
   \put(14.25,6.0){Bridge} 
   \put(24.0,6.0){Right lead} 
   \put(1.0,1.5){\dashbox{0.5}(11.5,4.0){~}} 
   \put(18.5,1.5){\dashbox{0.5}(11.5,4.0){~}} 
  \end{picture} 
 \end{center} 
\caption{\label{f:hel} 
  Schematic representation of the tight-binding model used to parameterize the
  Hamiltonian, Eq.\ (\protect\ref{e:defhs}).
 The circles depict the atomic sites for the leads and the molecular orbital
for the bridge (with energy written below) and
 the lines indicate the nonzero hopping amplitudes (written above).}
\end{figure}
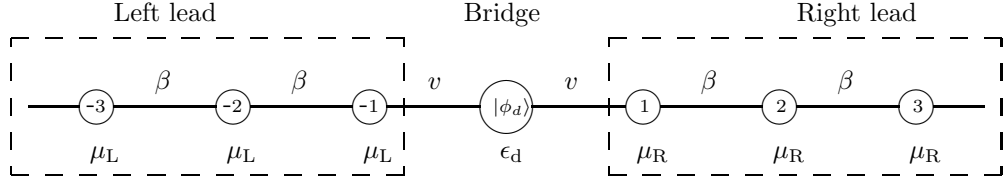

The electronic parameters of the model Hamiltonian (\ref{e:defhs}) can
in principle be determined by electronic
structure calculations \cite{bmd85}. In the model studies considered
below, we have adopted a parameterization which is based on a resonance
description of the molecular
bridge and a simple tight-binding model for the leads, schematically
shown in Fig.\ \ref{f:hel}. The molecular resonance is described by  the
discrete state $|\phi_{\rm d}\rangle$ and the states $|l \rangle$,
$l=\pm 1,\pm 2,...$ represent the atomic sites of the left ($-$) and
right ($+$) lead, respectively. It should be noted that in contrast to
the
lead states, the discrete state does not correspond to a single-site
tight-binding description (e.g. a single atomic orbital),
but rather is a molecular resonance state, and thus comprises
typically contributions from many atomic orbitals.
This molecular orbital can be constructed for molecule under bias
resulting in voltage-dependent $|\phi_{\rm d}\rangle$.
For information on the construction of molecular resonance states we
refer the reader to Ref. \cite{d91}.
  The nearest-neighbor coupling constants between two lead sites and
between the leads and the molecule are specified by  $\beta$ and $v$,
respectively, and $\mu_{\rm L/R}$ denotes the chemical potential
in the leads.
The stationary continuum states in the right lead  are given by
\begin{equation}
  |\phi_{k}\rangle =    
    \sum_l\frac{\sin(kl)}{\sqrt{\pi\beta\sin
k}}|l\rangle,
\end{equation}
and similar for the left lead. The energy $\epsilon$ of the electron
satisfies the dispersion relation
\begin{equation}\label{e:disp}
  \epsilon=\epsilon_{k\alpha}=\mu_{\alpha}+2\beta\cos k,
\end{equation}
in the left and right lead, $\alpha={\rm L/R}$, respectively.
Using this particular model for the leads we obtain 
\begin{equation}
   V_{{\rm{d}}k\alpha}\equiv
    \langle\phi_{\rm{d}}|H_{\rm{el}}|\phi_{k\alpha}\rangle =
    v\sqrt{\frac{\sin k}{\pi\beta}}.
\end{equation}

As we will see in Sec.\ \ref{sec:sol}, the electronic structure of the leads enters 
the expressions for the observables of interest only through the energy-dependent
width function of the leads
(atomic units with $e=\hbar=1$ are used throughout the paper unless stated otherwise)
\begin{equation}\label{e:gama}
  \Gamma_{\alpha}(\epsilon)
    \equiv 2\pi\sum_{k}\delta(\epsilon-\epsilon_{k\alpha})
           |V_{{\rm d}k\alpha}|^2.
\end{equation}
The width function $\Gamma_{\alpha}(\epsilon)$ is the imaginary part
of the self-energy  function\begin{equation} 
 \Sigma_{\alpha}(\epsilon)=\sum_{k}  
    \frac{|V_{dk\alpha}|^2}{\epsilon^{+}-\epsilon_{k\alpha}} 
 \equiv \Delta_{\alpha}(\epsilon)
       -{\textstyle\frac{i}{2}}\Gamma_{\alpha}(\epsilon),
\end{equation}
where $\epsilon^+=\epsilon + i\gamma$, $\gamma$ being a positive infinitesimal.
The real part of the self-energy function, the level-shift function 
$\Delta_{\alpha}(\epsilon)$,
is related to the width function via Hilbert transformation, i.e.\
\begin{equation}\label{e:htransform}
 \Delta_{\alpha}(\epsilon)=\frac{1}{2\pi} {\rm P} \int
   \frac{\Gamma_{\alpha}(\epsilon')}{\epsilon-\epsilon'}
 {\rm d}\epsilon',
\end{equation}
where ${\rm P}$ denotes the principal value of the integral.

For the nearest-neighbor tight-binding model of the leads introduced above,
the self-energy  function 
$\Sigma(z)$ is given by the Hubbard Green's function \cite{economou,ds},
multiplied by the coupling strength $v$ between the last 
atomic site in the leads and the bridge,
\begin{equation}
 \Sigma_{\alpha}(z) =
   \frac{2v^2}{z-\mu_{\alpha}+\sqrt{(z-\mu_{\alpha})^2-4\beta^2}}.
\end{equation}
Here, the width of the conduction band is given by $4\beta$. 
Analytic continuation in the complex energy ($z$) plane gives the real 
part
\begin{equation}
\Delta_{\alpha}(\epsilon)
=\left\{\begin{array}{ll}
      \frac{v^2}{2\beta^2}(\epsilon-\mu_{\alpha}) & 
      \qquad\mbox{for $|\epsilon-\mu_{\alpha}|<2\beta$}\\
      \frac{v^2}{2\beta^2}\left[
                             (\epsilon-\mu_{\alpha})
                             \mp\sqrt{(\epsilon-\mu_{\alpha})^2-4\beta^2}
                           \right] & 
      \qquad\mbox{for $\pm(\epsilon-\mu_{\alpha})>2\beta$}
    \end{array}
\right.
\end{equation}
and the imaginary part
\begin{equation}\label{e:tbgamamodel}
  \Gamma_{\alpha}(\epsilon)
     =
    \left\{\begin{array}{ll}
      \frac{v^2}{\beta^2}\sqrt{4\beta^2-(\epsilon-\mu_{\alpha})^2} & 
           \qquad\mbox{for $|\epsilon-\mu_{\alpha}|<2\beta$} \\
      0 & \qquad\mbox{for $|\epsilon-\mu_{\alpha}|>2\beta$}
    \end{array}
\right.
\end{equation}
As will be demonstrated in Sec.\ \ref{sec:results},
the inclusion of the 
energy dependence of the width function $\Gamma_{\alpha}(\epsilon)$ and thus of 
the coupling matrix elements $V_{{\rm{d}}k\alpha}$, which complicates
the theoretical treatment significantly, is crucial in order to
account correctly for inelastic effects, in particular for energies 
close to the edge of the conduction band. 

To study vibrationally inelastic effects on the transmission through the molecular bridge, 
we consider a single
vibrational (reaction) mode, along which the
equilibrium geometry of the discrete electronic state is shifted
with respect to the continuum states due to the presence of the 
additional electron at the molecule.
Within the harmonic approximation, this situation is described
by the  vibrational Hamiltonians
\begin{equation}\label{e:hvib}
  \tilde H_0=\omega_{\rm S}\, a^{\dagger}a 
  \quad , \quad
   \tilde H_d=\omega_{\rm S}\, a^{\dagger}a + \lambda (a+a^{\dagger}) + \epsilon_{\rm d}
     =\omega_{\rm S}\, a_{\rm d}^{\dagger}a_{\rm d}
        + \epsilon_{\rm d} - \frac{\lambda^2}{\omega_{\rm S}}.
\end{equation}
Here, $\omega_S$ is the vibrational frequency of the reaction mode, and
$a^{\dagger}$ and $a$ denote the creation and annihilation operators
for the reaction mode which are related to the corresponding operators in the discrete
electronic state by the shift in equilibrium geometry, $\lambda/(\sqrt{2}\omega_{\rm S})$ , 
i.e.\  $a_{\rm{d}} = a+\lambda/\omega_{\rm S}$.

Most of the experimental studies of electron transport through molecular bridges 
conducted  so far have considered relatively large molecules 
with many vibrational degrees of freedom. In large molecules
the coupling of reaction coordinates (which are strongly coupled to 
the electronic degrees of freedom)
to the remaining (inactive) vibrational modes of the molecule results in the
process of intramolecular vibrational redistribution, which is well known from the spectroscopy 
of large molecules \cite{Hopkins80,Mukamel80,Freed83,Brumer91,Uzer91,Zewail94,Nesbitt96}.
To study the effect of vibrational relaxation on the electron transmission,
we adopt a linear response model for vibrational relaxation \cite{td98} in the
discrete state, where the reaction mode is coupled to
a bath of harmonic oscillators. 
Thus the Hamiltonian of the overall system is given by  
\begin{equation}\label{e:defh}
  H=H_{\rm S} + H_{\rm B} + H_{\rm SB},
\end{equation}
where the 'system' Hamiltonian $H_{\rm S}$ is given by Eq.\ (\ref{e:defhs}),
the bath Hamiltonian reads
\begin{equation}\label{e:hbath}
  H_{\rm B}=\sum_{j} \omega_{j}\, b^{\dagger}_{j}b_{j},
\end{equation}
and the coupling between the reaction coordinate and the bath is
given by the bilinear interaction
\begin{equation}\label{e:defhsb}
H_{\rm SB}=|\phi_{\rm d}\rangle\sum_{j} c_{j}
           (a_{\rm d}b_{j}^{\dagger} + a_{\rm d}^{\dagger}b_{j})
           \langle\phi_{\rm d}|.
\end{equation}
Here, $b_j^{\dagger}$ and $b_j$ denote creation and annihilation operators for the bath
mode with frequency $\omega_j$ and $c_j$ is the corresponding system-bath coupling constant.
There is no bilinear coupling between the system and the bath modes if
the molecule is in the neutral state (and thus the electron in the lead states)
in our form of $H_{\rm SB}$. In reality there is,
of course, such coupling and also relaxation
due to anharmonic effects in the neutral state of the molecular bridge,
describing, for example
vibrational relaxation processes after an electron has been transmitted
through the bridge.
The study of these effects, which for the process of resonant
electron-molecule scattering
were found small compared to the relaxation mechanism in the discrete
molecular state \cite{td98}, will be the subject of future work.

As is well known \cite{Leggett87,Weiss99}, all properties of the vibrational bath
which influence the dynamics of the system are characterized by the bath spectral density
\begin{equation} 
J(\omega) =\sum_j c_j^2\delta(\omega-\omega_j).
\end{equation} 
In the numerical calculations reported below, it will be modeled
by a continuous Ohmic bath with exponential cutoff \cite{Leggett87,Weiss99}
\begin{equation}\label{e:spec}
J(\omega)=\eta \omega e^{-\omega/\omega_{\rm c}}.
\end{equation}
Here, the characteristic frequency $\omega_c$ defines the maximum of the spectral density
and the overall strength of the system-bath coupling is measured by $\eta$.

\subsection{Observables of Interest}

Various observables are of interest when investigating the influence of vibrational motion
on the electron transport trough a molecular bridge.
The most detailed information on the transmission process of a single
electron is comprised in the initial- and
final-state resolved scattering probability. 
Employing scattering theory, it is straightforward to show that
the  probability for scattering of
an electron with the energy $\epsilon_i$ from lead $\alpha_i$ into a
state with energy $\epsilon_f$ in the lead $\alpha_f$, accompanied by a
vibrational transition from state
$|v_i\rangle |{\bf v}_i\rangle$
to state $|v_f\rangle |{\bf v}_f\rangle$%
\footnote{The formula has to be modified for elastic backscattering, i.\ e.\ 
          $\alpha_i=\alpha_f$, 
          $|v_i\rangle |{\bf v}_i\rangle =|v_f\rangle |{\bf v}_f\rangle$,
          confront (\ref{e:totelb})}
(we denote the bath vibrations with bold face letters and the system
mode vibrations with italic letters)
is given by \cite{d91}
\begin{eqnarray}\label{e:t}
 t_{\alpha_f\leftarrow\alpha_i}
 (\epsilon_f,v_f,{\bf v}_f,\epsilon_i,v_i,{\bf v}_i)
&=&
 \delta (\epsilon_i+E_{{\bf v}_i}+E_{v_i}-\epsilon_f-E_{{\bf v}_f}-E_{v_f})
  \Gamma_{\alpha_i}(\epsilon_i)\Gamma_{\alpha_f}(\epsilon_f)
\nonumber\\ \label{tgen} 
  &&\qquad
  \times \left|
    \langle{\bf v}_f|\langle v_f|\langle\phi_{\rm d}|
         (\epsilon_i^+ -H)^{-1}
    |\phi_{\rm d}\rangle|v_i\rangle|{\bf v}_i\rangle
  \right|^2,
\end{eqnarray}
where  $E_{{\bf v}_i}$, $E_{v_i}$ and $E_{{\bf v}_f}$, $E_{v_f}$ are the initial and final
vibrational energies 
of the bath and system modes, respectively.  Writing the transition 
probability in the form (\ref{tgen}), we have assumed, for simplicity,
that the coupling element
$V_{{\rm d}k\alpha}$ does not depend on the vibrational degrees of freedom. The generalization 
of the formula to include such effects is straightforward (see, for example, Ref.\ \cite{d91}).

While Eq.\ (\ref{e:t}) describes the most detailed information on the 
scattering process, in experiments typically more averaged observables are measured.
It is thus expedient  to introduce the integral transmission probability from lead 
$\alpha_i$ into lead $\alpha_f$ (summed over all possible final vibrational
states)
\begin{equation}\label{e:tsum}
 t_{\alpha_f\leftarrow\alpha_i}(\epsilon_f,\epsilon_i)\equiv
 \sum_{v_f,{\bf v}_f}t_{\alpha_f\leftarrow\alpha_i}
 (\epsilon_f,v_f,{\bf v}_f,\epsilon_i,v_i,{\bf v}_i),
\end{equation}
and the total transmission probability, integrated over the final energy of the electron,
$\epsilon_f$,
\begin{equation}\label{e:ttot}
 t_{\alpha_f\leftarrow\alpha_i}(\epsilon_i)=
   \int t_{\alpha_f\leftarrow\alpha_i}(\epsilon_f,\epsilon_i) 
       \,{\rm d}{\epsilon_f}.
\end{equation}

The most important observable for the study of electron transport
trough a molecular bridge is, of course, the
current which is induced when a finite voltage is applied to the molecular junction. 
To calculate the current through the bridge, we employ
the generalized Landauer formula \cite{n01}
\begin{equation}\label{e:landauer}
 I=\frac{1}{\pi}\int{\rm d}\epsilon_i\int{\rm d}\epsilon_f
   \left\{
     t_{R\leftarrow L}(\epsilon_f,\epsilon_i)
     f_L(\epsilon_i)[1-f_R(\epsilon_f)]
    -
     t_{L\leftarrow R}(\epsilon_f,\epsilon_i)
     f_R(\epsilon_i)[1-f_L(\epsilon_f)]
   \right\},
\end{equation}
where $f_{\alpha}(E)$, $\alpha=L,R$, denotes the Fermi-Dirac distribution for the
left and right lead, respectively. 

In principle the basis states $|\phi_d\rangle$, $|\phi_{k\alpha}\rangle$, and 
therefore also the functions $\Gamma_{\alpha}(\epsilon)$ and vibrational 
Hamiltonian $H_d$, depend on the bias voltage $V$ across the bridge.
Here, we assume for simplicity that
the bias voltage $V$ enters Eq.\ (\ref{e:landauer}) only through the Fermi distribution 
of the leads 
and the width function $\Gamma_{\alpha}(\epsilon)$ [cf.\ Eq.\ (\ref{e:disp}), (\ref{e:gama})]
via the chemical potentials of the leads $\mu_{L/R} = \pm V/2$. 
Moreover, in this paper we will not consider thermal effects on the electron transport,
i.e.\ in all numerical calculations reported below we have taken $T=0$ K.
Thus, the initial state of the vibrational degrees of freedom
is the ground state of the system and bath modes, respectively,
$|v_i=0\rangle |{\bf v}_i={\bf 0}_B\rangle$ and the Fermi-Dirac distribution 
$f_{\alpha}(E)$ in Eq.\ (\ref{e:landauer})
reduces to the Heaviside step function. 

The validity of the formula (\ref{e:landauer}) 
for calculating currents including inelastic effects
has been discussed controversially in the literature. The current formula 
is sometimes used without the Pauli exclusion principle factors
$[1-f(\epsilon)]$ (see for example the discussion in Refs.\ \cite{datta,w00b,s92}). 
Employing nonequilibrium
Greens-function theory and the second quantized version of the
Hamiltonian (\ref{e:defhs}), it can be shown that Eq.\ (\ref{e:landauer}) gives 
the correct description
of the current in the limit
where many-electron processes are negligible for the dynamics \cite{Cizekunp}.
This limit is sometimes called the single particle approximation. For the study
of the tunnelling through a vibrating single-molecule junction beyond this approximation
(but within the wide-band limit) see \cite{f04} and references therein.
It should also be mentioned that the use
of Eq.\ (\ref{e:landauer}) implicitly assumes that  the bridge relaxes 
into the ground vibrational state
$|v_i=0\rangle |{\bf v}_i={\bf 0}_B\rangle$ before each
subsequent transmission event. In view of the low currents observed experimentally, this should be
a reasonable assumption (for a similar discussion in the case of STM currents,
see, for example, Ref.\ \cite{ga93}).

\subsection{Method of Solution}\label{sec:sol}

We explain the method of solution of the problem in this paragraph, starting
from simplified cases which are later also discussed and shown for comparison 
with the final solution.

\subsubsection{Elastic transmission}

Let us first consider the case without coupling to the vibrational degrees of freedom 
[i.e. $\lambda=\eta=0$ in Eqs.\ (\ref{e:hvib}),(\ref{e:spec})]. 
In this case, only elastic processes contribute to the electron transport,
and the total transmission probability is given by
\begin{equation}\label{e:tel}
 t_{R\leftarrow L}(\epsilon_i)
= t_{L\leftarrow R}(\epsilon_i)
  =\frac{\Gamma_{R}(\epsilon_i)\Gamma_{L}(\epsilon_i)}{
    [\epsilon_i-\epsilon_{\rm d}-\Delta_L(\epsilon_i)-\Delta_R(\epsilon_i)]^2
       +[\Gamma_L(\epsilon_i)+\Gamma_R(\epsilon_i)]^2/4},
\end{equation}
which is the well-known result for elastic resonant tunneling transmission.
The Landauer formula (\ref{e:landauer}) for the current is exact 
in this case (see, for example, Ref.\ \cite{haug-jauho})
and the evaluation of the current reduces to a simple numerical
integration.


\subsubsection{Inelastic transmission without dissipation}

Let us next consider the influence of the coupling to the vibrational degrees of freedom
on the transmission of the electron, i.e.\ vibrationally inelastic transmission through
the molecular bridge. If we exclude dissipative processes induced by the coupling to the bath,
the dynamics is described by the system Hamiltonian $H_S$, Eq.\ (\ref{e:defhs}).
The probability for transmitting an electron  with energy $\epsilon_i$
from the left lead into  a state with energy $\epsilon_f$ in the right
lead, accompanied by a vibrational transition from the state
$|v_i=0\rangle$ (which is the initial vibrational state at temperature
$T=0$) to the state $|v_f\rangle$ is given by (\cite{s92,nf02})
\begin{equation}\label{e:tvib}
 t_{\rm R\leftarrow L}(\epsilon_f,\epsilon_i) =
   \sum_{v_f}\delta(\epsilon_i-\epsilon_f-E_{v_f})
   \Gamma_{\rm R}(\epsilon_f)\Gamma_{\rm L}(\epsilon_i)
   \left|\langle{v}_f|G_{\rm d}^{(S)}(\epsilon_i)|0\rangle\right|^2 ,
\end{equation}
where  
\begin{equation}\label{e:defgds}
  G_d^S(E)\equiv
  \langle\phi_{\rm d}|(E^+-H_S)^{-1}|\phi_{\rm d}\rangle\
\end{equation}
denotes the Green's function projected on the resonance state. The exact one-particle 
transmission probability (\ref{e:tvib}) can be obtained from the solution of both 
time-dependent \cite{bhg93,gsl89} and time-independent Schr\"{o}dinger equation (see for example 
\cite{shls91,hb99,s92,nf02}). 
The Green's function (\ref{e:defgds}) can also be written in closed form. 
Employing projection operator techniques well known from  the theory of resonant electron-molecule
scattering (see \cite{d91} and references therein),
$G_d^S(E)$ can be recast in the form
\begin{equation}\label{e:defgds2}
  G_d^S(E)
  =\left[ E^+-\tilde H_d-\Sigma_L(E-\tilde H_0)-\Sigma_R(E-\tilde H_0)\right]^{-1}.
\end{equation}
This form has the advantage that the electronic continuum has been formally eliminated
and only vibrational dynamics in the discrete electronic space
has to be evaluated. We would like to stress that we do not assume 
the wide-band approximation (see the following section) 
in Eq. (\ref{e:defgds2}) and the Hamiltonian
operator $H_0$ enters the energy dependence of $\Sigma$.
Efficient techniques have been developed \cite{h95,chm00}
to evaluate the matrix elements of the Green's function (\ref{e:defgds2}).
In the present case, where $\tilde H_0$ and $\tilde H_d$ describe harmonic oscillators,
the  Green's function (\ref{e:defgds2}) can be obtained, e.g., by inverting a basis
representation of the operator $E^+-\tilde H_d-\Sigma_L(E-\tilde H_0)-\Sigma_R(E-\tilde H_0)$
for each energy $E$, employing efficient algorithms 
for the inversion of tridiagonal  matrices \cite{dc80}.

\subsubsection{Wide-band approximation}

While the expressions discussed so far take full account of the energy dependence
of the width function $\Gamma(\epsilon)$, in the majority of previous work
on the effect of vibrational motion on  electron transmission, the so-called wide-band 
(WB) approximation  has been invoked, where the width function is assumed to be constant, 
i.e.\ $\Gamma(\epsilon)=const$
(for recent examples see \cite{ab03,bf04}). 
In this approximation, the level-shift function
vanishes [cf.\ Eq.\ (\ref{e:htransform})], $\Delta(\epsilon)=0$. Introducing the 
eigenstates $|n\rangle$ and eigenenergies $E_n$ for the operator $\tilde H_d$ the   
transmission probability, Eq.\  (\ref{e:tvib}), can be written as 
\begin{equation}
 t_{\rm R\leftarrow L}^{({\rm WB})}(\epsilon_f,\epsilon_i) =
   \Gamma_{\rm R}\Gamma_{\rm L}
   \sum_{v_f}\delta(\epsilon_i-\epsilon_f-E_{v_f})
   \left|
   \sum_{n}\frac{\langle v_f|n\rangle\langle n|0\rangle
     }{\epsilon_i-E_n+\frac{i}{2}\,(\Gamma_{\rm R} + \Gamma_{\rm L})}
   \right|^2,
\end{equation}
which yields after integration over the final energy of the electron
the total transmission probability
\begin{eqnarray}\label{e:tWB}
 t_{\rm R\leftarrow L}^{({\rm WB})}(\epsilon_i) 
 &=&\Gamma_{\rm R}\Gamma_{\rm L}
  \sum_{n}\frac{|\langle{n}|0\rangle|^2}{
                (\epsilon_i-E_n)^2+\frac{1}{4}\,(\Gamma_{\rm R} + \Gamma_{\rm L})^2} \\
 &=&\Gamma_{\rm R}\Gamma_{\rm L}
  \sum_n\frac{\frac{1}{n!}\left(\frac{\lambda^2}{2\omega_{\rm S}^2}\right)^n
              e^{-\frac{\lambda^2}{2\omega_{\rm S}^2}}
              }{(\epsilon_i-E_n)^2+\frac{1}{4}\,(\Gamma_{\rm R} + \Gamma_{\rm L})^2}. \nonumber
\end{eqnarray}
In Sec.\ \ref{sec:results}, we will study the validity 
of the wide-band approximation based on the comparison
of Eq.\ (\ref{e:tWB}) 
with the full inelastic transmission probability given by Eq.\ (\ref{e:tvib}).


\subsubsection{Inelastic transmission including vibrational relaxation}

Finally, we consider the case of inelastic transmission of an  electron through the molecular bridge
in the presence
of vibrational relaxation.
The theoretical treatment of this problem is considerably more complicated than the cases
considered above, because we have to deal with two qualitatively different continua:
the electronic scattering continuum describing the leads and the dissipative vibrational 
mode continuum of the bath.
In principle, it is possible to start from the general formula for the 
transition probability (\ref{tgen})
and  derive  formal expressions  as in Eqs.\ (\ref{e:tvib}), (\ref{e:defgds2}) 
with the bath modes included  in $\tilde H_0$ and $\tilde H_d$. However, 
with increasing number of bath modes it becomes
difficult to perform the matrix inversion in Eq.\ (\ref{e:defgds2})
 and, therefore, such an approach
is limited to very few bath modes. 

To circumvent this problem,
we adopt an approach that has been proposed  to describe the effect of vibrational
relaxation in the context of resonant electron scattering from large molecules \cite{td98}.
The basic idea of this method is to express the transmission probability, Eq.\ (\ref{e:tsum}),
as a sum
\begin{eqnarray}\label{tser}
 t_{\rm R\leftarrow L}(\epsilon_f,\epsilon_i) &=&
    \sum_{m=0}^{\infty} t_{\rm R\leftarrow L}^{(m)}(\epsilon_f,\epsilon_i),
\end{eqnarray}
where $t_{\rm R\leftarrow L}^{(m)}(\epsilon_f,\epsilon_i)$ describes
transmission processes with $m$ quanta
of excitation in the final state of the bath (as was mentioned above,
we assume that the initial state of the bath at temperature $T=0$
is given by  $|{\bf v}_i={\bf 0}_B\rangle$).
In the parameter regime where the vibrational relaxation rate is small compared
to the electronic decay rate of the resonance, only the first few terms in 
the expansion will contribute to the overall transmission probability.
The first two terms in the expansion, which correspond to processes where the final state of the 
bath contains
zero or one quantum of excitation, respectively, read
\begin{mathletters}\label{tserall}
\begin{eqnarray}
\label{e:t0}
 t_{\rm R\leftarrow L}^{(0)}(\epsilon_f,\epsilon_i) &=&
   \sum_{v_f}\delta(\epsilon_i-\epsilon_f-E_{v_f})
   \Gamma_{\rm R}(\epsilon_f)\Gamma_{\rm L}(\epsilon_i)
   \left|\langle{v}_f|G_{\rm d}(\epsilon_i)|0\rangle\right|^2,\\
\label{e:t1}
 t_{\rm R\leftarrow L}^{(1)}(\epsilon_f,\epsilon_i) &=&  
   \sum_{v_f}J(\epsilon_i-\epsilon_f-E_{v_f})
   \Gamma_{\rm R}(\epsilon_f)\Gamma_{\rm L}(\epsilon_i)
   \left|\langle{v}_f|
     \,G_{\rm d}(\epsilon_f+E_{v_f})\,a_{\rm d}\,G_{\rm d}(\epsilon_i)\,
   |0\rangle\right|^2,
\end{eqnarray}
\end{mathletters}
with the Green's function
\begin{eqnarray}\label{gd}
 G_{\rm d}(E)& \equiv & \langle\phi_d|\left( E^+ - H_{\rm S} -
    |\phi_d\rangle a^{\dagger}_d\langle\phi_d|
    \int {\rm d}\omega\frac{J(\omega)}{E^+-H_{\rm S}-\omega}
    |\phi_d\rangle a_d\langle \phi_d|
    \right)^{-1}|\phi_d\rangle \\
 &=& \langle\phi_d|\left( E^+ - H_{\rm S} -
    |\phi_d\rangle a^{\dagger}_d
    \int {\rm d}\omega \: J(\omega) \: G_d^S(E-\omega)
     \: a_d\langle \phi_d|
    \right)^{-1}|\phi_d\rangle . \nonumber
\end{eqnarray}
The expressions for higher order terms ($m>1$) can be found in Ref.\ \cite{td98}. 
The Green's function $G_{\rm d}(E)$ can be evaluated 
either as the unique solution of the Lippmann-Schwinger equation 
\begin{equation} 
 G_{\rm d}(E)=G_{\rm d}^S(E)+G_{\rm d}^S(E) 
   \left[a^{\dagger}_d 
     \int {\rm d}{\omega} \: J(\omega)\: G_{\rm d}^S(E-\omega) 
   \,a_d\right]\,G_{\rm d}(E) 
\end{equation} 
using the iterative Schwinger-Lanczos method \cite{chm00,mhc91} (which converges 
with only few iterations) or employing matrix inversion techniques as described in 
Ref.\ \cite{td98}.

In general situations where the left and right lead are not identical, e.g., due to a 
non-zero bias voltage, 
the expansion in Eq.\ (\ref{tser}) has to be terminated for practical reasons 
at a certain order $m$ and is thus only applicable in the case of a weakly damped system mode.
We would like to emphasize, though, that the electronic coupling of the 
molecule to the leads is treated exactly in the approach outlined above and it 
can thus be arbitrarily strong. Furthermore, the treatment of the dissipation described by Eq.\ 
(\ref{tserall}) does not invoke  Markov-type approximations and is, therefore, not limited
to situations where the bath-correlation time is short compared to the system dynamics. 

In the case of identical left and right leads
and zero bias, however, one can exploit the unitarity condition 
to express the sum of all higher order
corrections, i.e.\  $t_{R\leftarrow L}^{(m)}$ with $m>0$, in terms of the elastic (with
respect to the bath) terms $t_{R\leftarrow
L}^{(0)}$ and{\label{ftunit}} $t_{L\leftarrow L}^{(0)}$. To see
this, it is noted that in the symmetric case we have
$t_{R\leftarrow L}^{(m)}=t_{L\leftarrow L}^{(m)}$ for all $m>0$. 
For m=0, the total transmission probabilities are given by
\begin{mathletters}
\begin{eqnarray}\label{e:totel}
 t_{\rm R\leftarrow L}^{(0)}(\epsilon_i)&=& 
   \sum_{v_f=0}^{\infty} 
   \Gamma_{\rm R}(\epsilon_i-E_{v_f})\Gamma_{\rm L}(\epsilon_i)
   \left|\langle v_f |G_{\rm d}(\epsilon_i)|0\rangle
   \right|^2 ,
   \\
 \label{e:totelb}
 t_{\rm L\leftarrow L}^{(0)}(\epsilon_i) &=& 
   \left|
     1-i\Gamma_{\rm L}(\epsilon_i)\langle{0}|G_{\rm d}(\epsilon_i)|0\rangle\right|^2
     +\sum_{v_f=1}^{\infty} \Gamma_{\rm L}(\epsilon_i-E_{v_f})\Gamma_{\rm L}(\epsilon_i)
   \left|\langle{v_f}|G_{\rm d}(\epsilon_i)|0\rangle
   \right|^2.
\end{eqnarray}
\end{mathletters}
The difference between the transmissions probabilities from the left to the right lead
and from the left back to the left lead is due to 
the special role of the transition amplitude with the same initial and final state
which only contributes in the latter case. 
Together with the unitarity condition 
\begin{equation}
 t_{\rm L\leftarrow L}^{(0)}(\epsilon)
 +\sum_{m>0} t_{\rm L\leftarrow L}^{(m)}(\epsilon)
 +t_{\rm R\leftarrow L}^{(0)}(\epsilon)
 +\sum_{m>0} t_{\rm R\leftarrow L}^{(m)}(\epsilon)=1 ,
\end{equation}
this yields
\begin{equation}\label{e:tunit}
   t_{R\leftarrow L}(\epsilon_i)
    = {\textstyle\frac{1}{2}}\left(
        1+t_{\rm R\leftarrow L}^{(0)}(\epsilon_i)
         -t_{\rm L\leftarrow L}^{(0)}(\epsilon_i)
                             \right)
    = -\Gamma_{\rm L}(\epsilon_i)\,
        {\rm Im}\langle 0|G_{\rm d}(\epsilon_i)|0\rangle.
\end{equation}
Eq.\ (\ref{e:tunit}) is an exact 
formula for the total transmission of the 
electron through the molecular bridge, 
including the bath to all orders in the system-bath coupling.
Moreover, it can easily be evaluated numerically.
Although it is limited to symmetric leads, and thus cannot be applied directly to calculate the current,
it is very helpful for checking the convergence 
properties of the expansion (\ref{tser}) for zero bias or in the linear 
response regime.


\section{Results and Discussion}\label{sec:results}

In this section we present the results of a model study of the influence of
vibrational motion and vibrational dissipation 
on the transmission probability and the current-voltage
characteristics of a molecular bridge.
To obtain a comprehensive picture of the  various mechanisms, 
we shall consider models in different parameter regimes.
In particular, we will consider both the case of a molecular bridge that
is weakly coupled to the leads, resulting in narrow resonance structures, and the 
opposite case of a broad resonance, caused by strong interaction with the leads. 
The parameters specific to the different models  
are collected in Table \ref{t:modpar}. The nearest-neighbor coupling strength 
in the leads is 
$\beta=1$ eV in all models considered (this parameter is an overall energy scaling factor).
The characteristic frequency of the bath is chosen to coincide with the frequency
of the system mode, i.e.\ $\omega_C=\omega_S$, and we will consider a relatively weak
coupling between the system mode and the bath, $\eta=0.1$.


\subsection{Tunneling through a narrow resonance}

First, we consider the resonant transmission of electrons in the 
tunneling regime, which is characterized by a relatively weak coupling
of the electronic state localized at the bridge to those in the leads.
We have chosen a coupling strength  of $v=0.2$ eV, which corresponds to 
a fifth of the nearest-neighbor
hopping amplitude $\beta$ in the leads. We will, furthermore, consider 
a localized state that is situated well inside the
conduction band with an energy of $\epsilon_d=0.5$ eV (for zero voltage,
the conduction band extends over the range [ -2 eV, 2 eV]). 

\begin{figure}
 \begin{center}
   \epsfxsize=0.49\textwidth \epsfbox{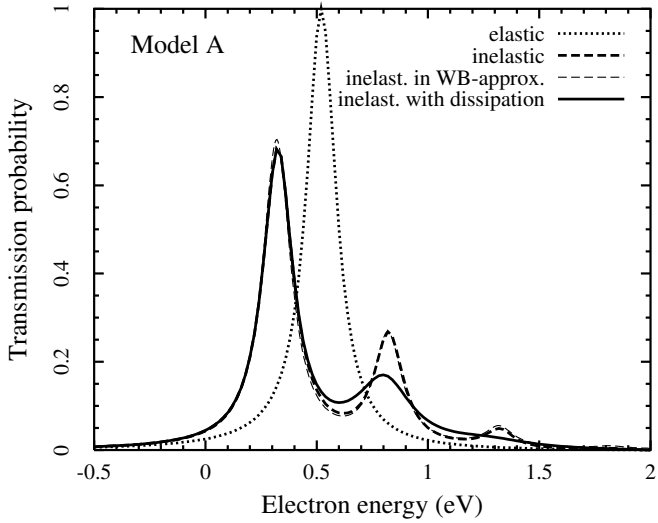}
 \end{center}
 \caption{\label{f:TvibF1}
   Transmission probabilities for Model A 
   at zero bias voltage. The results shown have been obtained at different levels of theory:
   purely elastic transmittance (dotted line), numerically exact inelastic transmittance (thick dashed line),
   inelastic transmittance  in the wide-band approximation (thin dashed line). The full line
   depicts the inelastic transmission probability  including vibrational relaxation.}
\end{figure}

We  start with Model A, which is characterized by a relatively 
weak coupling between the  electronic degrees of freedom and the  vibrational motion of the system mode, 
$\lambda=0.3$ eV. Fig.\ \ref{f:TvibF1} depicts the transmission probability 
for model A obtained for zero voltage
using  the different levels of theory introduced in Sec.\ \ref{sec:sol}: 
the elastic transmission probability, Eq.\  (\ref{e:tel}), 
the vibrationally inelastic transmission probability
integrated over the final electron energy, Eq.\ (\ref{e:tvib}), 
the wide-band approximation 
of the vibrationally inelastic transmission probability, Eq.\  (\ref{e:tWB}), 
and the transmission probability
in the presence of vibrational relaxation, Eq.\ (\ref{e:tunit}).
The elastic transmission probability (dotted line) exhibits a rather narrow peak at 
the position of the discrete electronic state.
Including the coupling of the electronic degrees of freedom to the vibrational motion of
system mode (vibrationally inelastic transmission, thick dashed line), 
this peak is seen  to become split into several sub-peaks which correspond to the different
vibrational levels in the discrete electronic state. This effect
is well known from previous studies \cite{wjw88,wjw89}.  
The comparison between the elastic and inelastic transmission
probability in Fig.\ \ref{f:TvibF1} demonstrates that even in the case of relatively weak 
electronic-vibrational coupling  a theoretical treatment which only includes elastic processes 
provides a rather poor description.
The wide-band approximation (thin dashed line), on the other hand, 
which takes the vibrational excitation into account
but neglects the energy dependence of the width function, is seen to give an excellent description
of the transmission probability. This is due to the fact that in model A the resonance
is situated well inside the conduction band and, therefore, threshold effects are 
are negligible.

The inclusion of the coupling to the vibrational bath (solid line)
causes a  further broadening and slight shift of the peaks belonging to the
first and second excited vibrational level in the discrete electronic state. 
The main peak, which corresponds to tunneling trough the ground vibrational state
of  $\tilde H_d$, on the other hand,
remains almost unaffected. This is due to the fact that 
(in a zero-order picture without electronic coupling)
the ground vibrational state of $\tilde H_d$ is a stationary state, while 
all excited vibrational states decay into the ground state. 

\begin{figure}
 \begin{center}
   \epsfxsize=0.49\textwidth \epsfbox{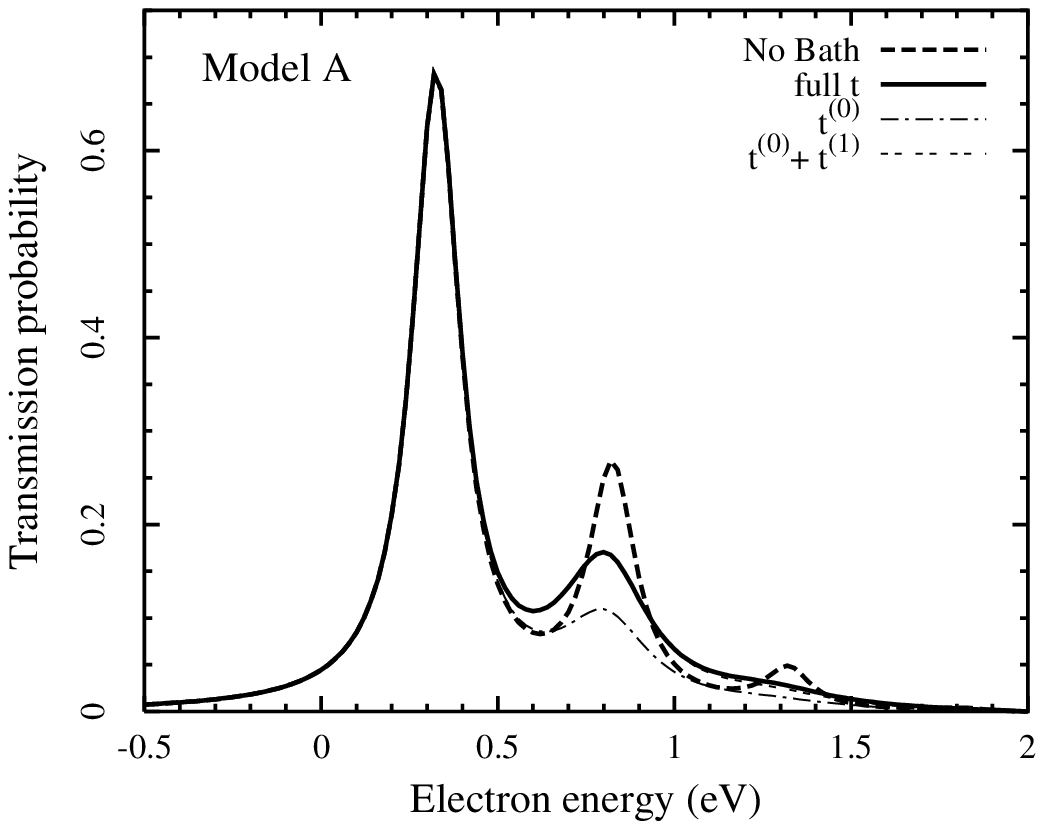}
 \end{center}
 \caption{\label{f:TbathF1}
   Transmission probability for Model A as in Fig.\ \ref{f:TvibF1}. 
   The different levels of approximation for the treatment of bath are 
   shown together with exact result (solid line) and the result obtained without
   coupling to the bath (thick dashed line). The result including up to one quantum in the
final state of the bath ($t^{(0)}+t^{(1)})$ is indistinguishable from the exact result
(full $t$).}
\end{figure}

The result for the inelastic electron transmission including 
vibrational relaxation depicted in 
Fig.\ \ref{f:TvibF1} has been  obtained employing Eq.\ (\ref{e:tunit}) which is only valid for zero bias.
For situations with non-zero bias (in particular, to evaluate the current through the 
bridge), 
we will use the expansion of the total transmission probability 
in terms of the number of excitations in the final states
of the bath, Eq.\ (\ref{tser}). It is therefore important 
to study the validity of this expansion. The results in Fig.\  \ref{f:TbathF1}
demonstrate that  for the present example the expansion (\ref{tser}) is well converged if bath states
with zero and one quantum of excitation ($m=0,1$) are taken into account.

\begin{figure}
 \begin{center}
   \epsfxsize=0.49\textwidth \epsfbox{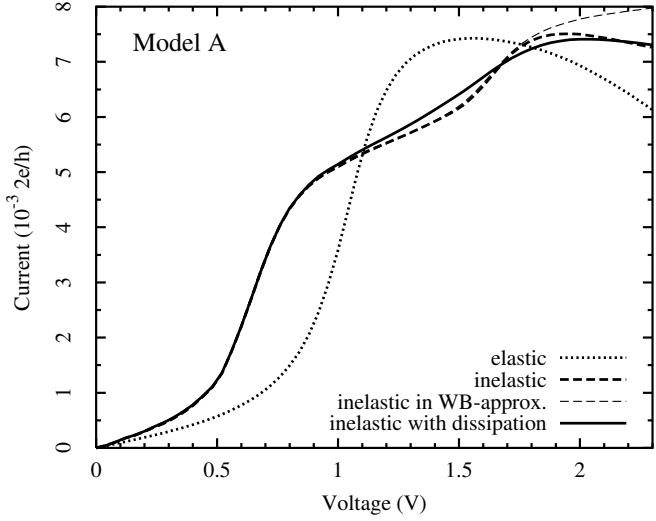}
 \end{center}
 \caption{\label{f:IVF1}
   Current-voltage characteristic for Model A obtained at different level of theory.}
\end{figure}

Fig.~\ref{f:IVF1} shows the current through the bridge as a function of the applied voltage
for model A. The results have been  obtained employing Eq.\ (\ref{e:landauer}).
As has been discussed in detail by other authors \cite{bp88,gs88}, each resonance peak in the
transmission probability has its 
counterpart in a step in the current-voltage curve. Thereby, the  steps occur in the order 
as they appear in the transmission probability counted from the zero-voltage Fermi energy 
(which is set to zero).
Accordingly, the  current based on the elastic treatment of the transmission (dotted line)
exhibits only a single broad maximum corresponding to the position of the discrete electronic state.
The inelastic current (thick dashed line), on the other hand, exhibits several steps which belong to the 
different vibrational peaks in the transmission probability. The   
wide-band approximation (thin dashed line) is seen to give very good results except at high voltages,
where the resonances are closer to the edge of the conduction band and, therefore, 
the energy dependence of the width function becomes important.

Although there is a pronounced effect of the vibrational motion on the current 
in model A, the influence of the coupling to the bath (thick solid line) is rather small. This is 
a consequence of the location of the discrete electronic state,
 which is situated 0.5 eV above 
the Fermi energy in this model.
As a result, the current for low voltages ($0 - 1$V) is almost exclusively
due to tunneling of electrons via the resonance corresponding to the ground state of $\tilde H_d$,
which is  hardly affected by dissipation (cf.\ the discussion above).

\begin{figure}
 \begin{center}
   \epsfxsize=0.49\textwidth \epsfbox{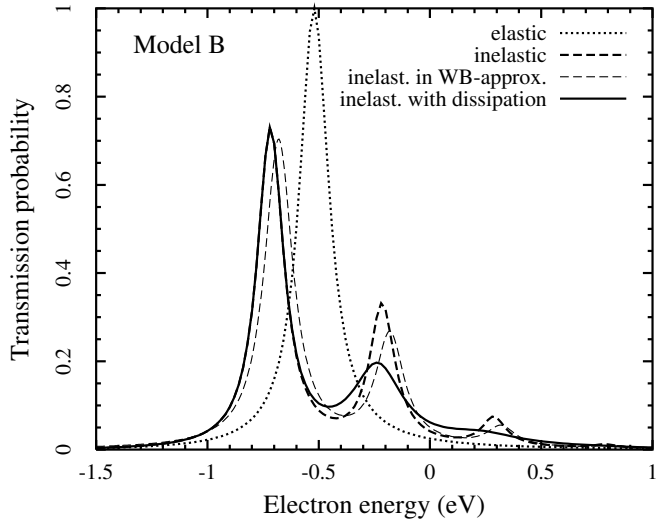}
 \end{center}
 \caption{\label{f:TvibF2}
   Transmission probabilities for Model B
   at zero bias voltage. Shown are results obtained at different levels of theory 
   as explained in the caption of Fig.\ \protect\ref{f:TvibF1}.}
\end{figure}

\begin{figure}
 \begin{center}
   \epsfxsize=0.49\textwidth \epsfbox{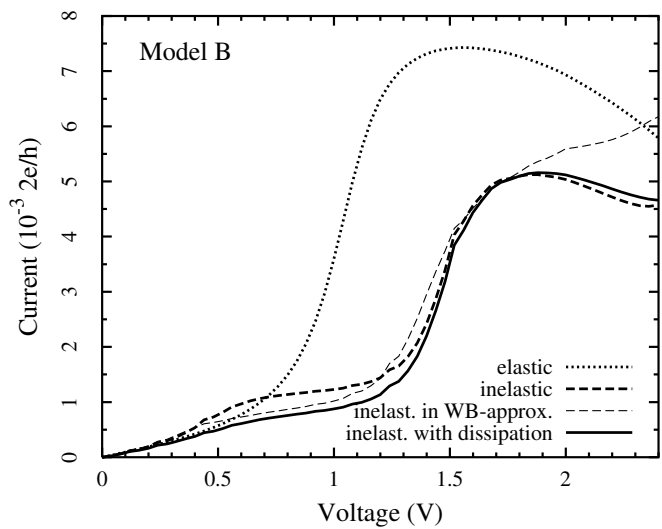}
 \end{center}
 \caption{\label{f:IVF2}
   Current-voltage characteristic for Model B. Shown are results obtained at different levels of theory.}
\end{figure}

If we  change the discrete state energy to $\epsilon_d=-0.5$ eV (Model B),  
the transmission functions, 
depicted in Fig.~\ref{f:TvibF2}, 
remain virtually unchanged except for a shift 
in energy by 1 eV. In contrast, the current-voltage characteristic for  
Model B, shown in Fig.\ \ref{f:IVF2}, is qualitatively different from that of Model A
(cf.\ Fig.~\ref{f:IVF1}).  
The reason is that the order of the peaks as counted from the Fermi energy 
is reversed and therefore the low-voltage region of the current is influenced by tunneling trough 
excited vibrational states of $\tilde H_d$ which are more strongly affected by the presence 
of the bath. The current with and without vibrational relaxation 
thus differs by more then 50 $\%$ for voltages in the range $0.5-1$ V.

\begin{figure}
 \begin{center}
   \epsfxsize=0.49\textwidth \epsfbox{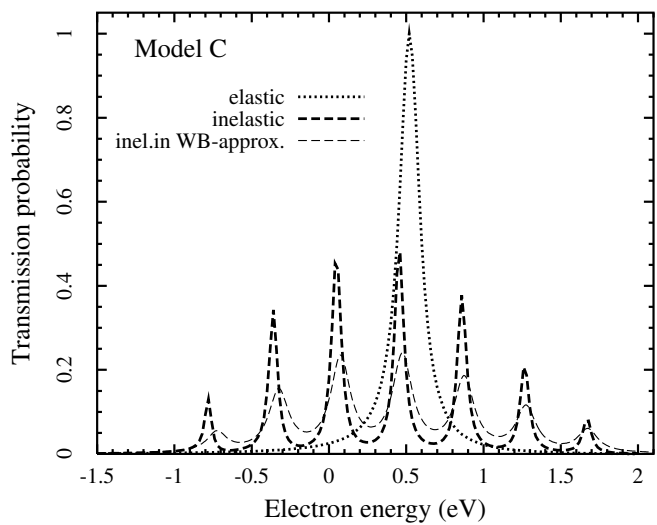}
 \end{center}
 \caption{\label{f:TvibXd}
   Transmission probabilities for Model C at zero bias voltage. 
   Shown are results obtained at different levels of theory.}
 \end{figure}

The differences between the various levels of the theoretical treatment
become more significant if we consider a model with larger coupling
between the electronic and vibrational degrees of freedom. 
Such a situation is described by model C, where the vibrational frequency is chosen
as $\omega_{\rm S}=0.4$ eV and the vibronic coupling strength as $\lambda=0.7$ eV. The location of the 
discrete state is the same as in model A,  $\epsilon_d=0.5$ eV. The transmission probability for
model C is shown in  Fig.~\ref{f:TvibXd}. It is seen that the stronger vibronic coupling
results in  a pronounced vibrational progression in the transmission probability.
In contrast to the cases considered above, the wide-band approximation 
essentially fails to describe the transmission probability.
It predicts an amplitude which is too small by about a factor of two. Furthermore,
the position of the peaks is not correctly described in 
the wide-band approximation due to the neglect of the  level-shift function 
$\Delta(E)$, and the individual peaks
in the numerically exact results are narrower than in the wide-band approximation. 
This effect of 'vibrational narrowing' is 
well-known from resonant electron-molecule scattering \cite{dc80} and results from the interference 
between overlapping resonances.

\begin{figure}
 \begin{center}
   \epsfxsize=0.49\textwidth \epsfbox{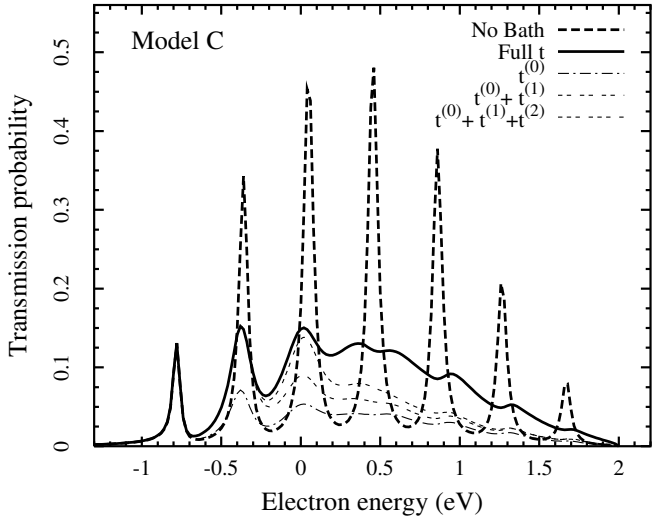}
 \end{center}
 \caption{\label{f:TbathXd}
   Transmission probabilities for Model C at zero bias voltage. Shown are the numerically
exact results for the inelastic transmittance with (thick dashed line) and without (full line)
vibrational relaxation, as well as the convergence of the expansion with respect to 
   the number of quanta in the final state of the bath  (thin lines).}
\end{figure}

The effect of vibrational relaxation on the transmission probability, 
illustrated in Fig.~\ref{f:TbathXd}, also is much more pronounced in
this model than in model A.
Except for the lowest two peaks, the  vibrational resonances are smeared into a broad hump,
when vibrational dissipation is included.
This is due to the fact that the vibrational relaxation process becomes more effective for higher excited
vibrational states. 
Also shown in Fig.~\ref{f:TbathXd} is the expansion of the total transmission probability in terms of
the number of excitations in the final state of the bath, Eq.\  (\ref{tser}).
As a result of the importance of higher vibrational states and the relatively small electronic coupling
to the leads, the expansion is seen to converge much slower than in the models considered above.
 
\begin{figure}
 \begin{center}
   \epsfxsize=0.49\textwidth \epsfbox{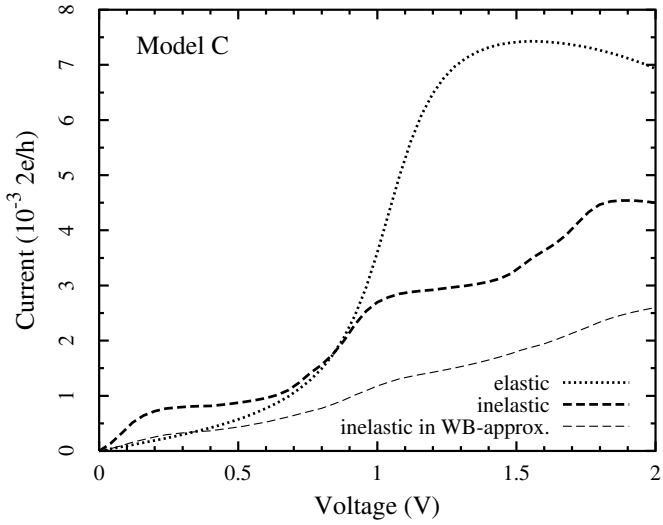}
 \end{center}
 \caption{\label{f:IVXd}
   Current-voltage characteristic for Model C. Shown are results obtained at different levels of theory 
    as explained in the legend.}
\end{figure}

\begin{figure}
 \begin{center}
   \epsfxsize=0.49\textwidth \epsfbox{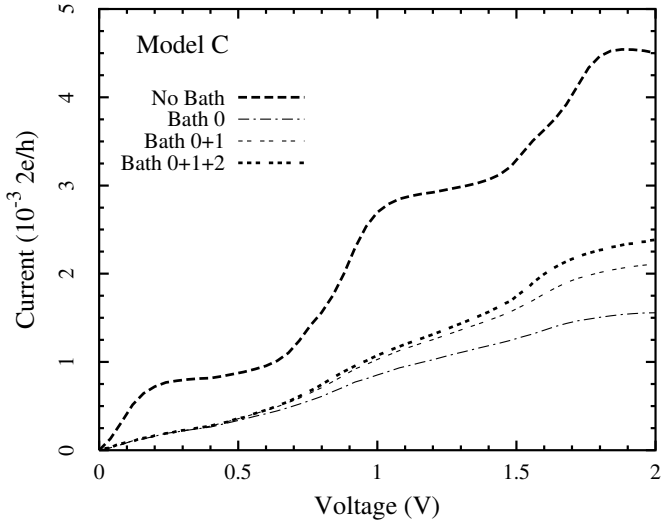}
 \end{center}
 \caption{\label{f:IVBathXd}
   Current-voltage characteristic for Model C. Shown are results that illustrate the 
  convergence of the expansion with respect to 
   the number of quanta in the final state of the bath, as well as the result without
   vibrational relaxation (thick dashed line)}
\end{figure}

Let us next consider the current-voltage characteristic for model C, depicted in Fig.\ \ref{f:IVXd}.
It is seen that the  coupling to the vibrational motion has a rather strong effect on the 
current through the bridge. In particular, the current-voltage characteristic
exhibits step-like structures corresponding to the different vibrational levels
in the discrete electronic state.
As expected from the discussion of the transmission probability above, both the elastic treatment and the 
wide-band approximation fail severely in the description of the current.
Fig.\ \ref{f:IVBathXd} demonstrates that 
the coupling to the bath has a strong effect on the 
current-voltage characteristic in model C. In particular, the step-like structure
is washed-out and the magnitude of the current decreases by more than a factor of two.
Also shown in Fig.\ \ref{f:IVBathXd} is the contribution of the different terms in the
expansion (\ref{tser}) to the total current.
In contrast to the transmission probability, the expansion for the current converges relatively fast
for this model. The reason for this at first sight surprising finding is that,
due to the Pauli principle exclusion factors in the formula for the current,
terms with a higher number of excitations in the final state of the bath
are suppressed at lower voltage.


\subsection{Transmission through a bridge strongly coupled to the leads}

All models considered so far were characterized by a relatively weak
coupling of the discrete electronic state to the leads, which results in narrow resonance structures.
In this section we shall consider the opposite limit of a localized state that
is strongly coupled to the leads.
Such a situation is realized in model D, where the coupling strength
between the discrete electronic state and the leads
is chosen as  $v=1$ eV. The position of the discrete state is in the upper part
of the conduction band,  $\epsilon_d=1.6$ eV. All other
parameters are the same as in Model C. 
We mention that if  the coupling to the bath is not considered, model D 
is essentially equivalent to a model studied 
by Gelfand et al. \cite{gsl89} in the context of 
inelastic tunneling in heterostructures.

\begin{figure}
 \begin{center}
   \epsfxsize=0.49\textwidth \epsfbox{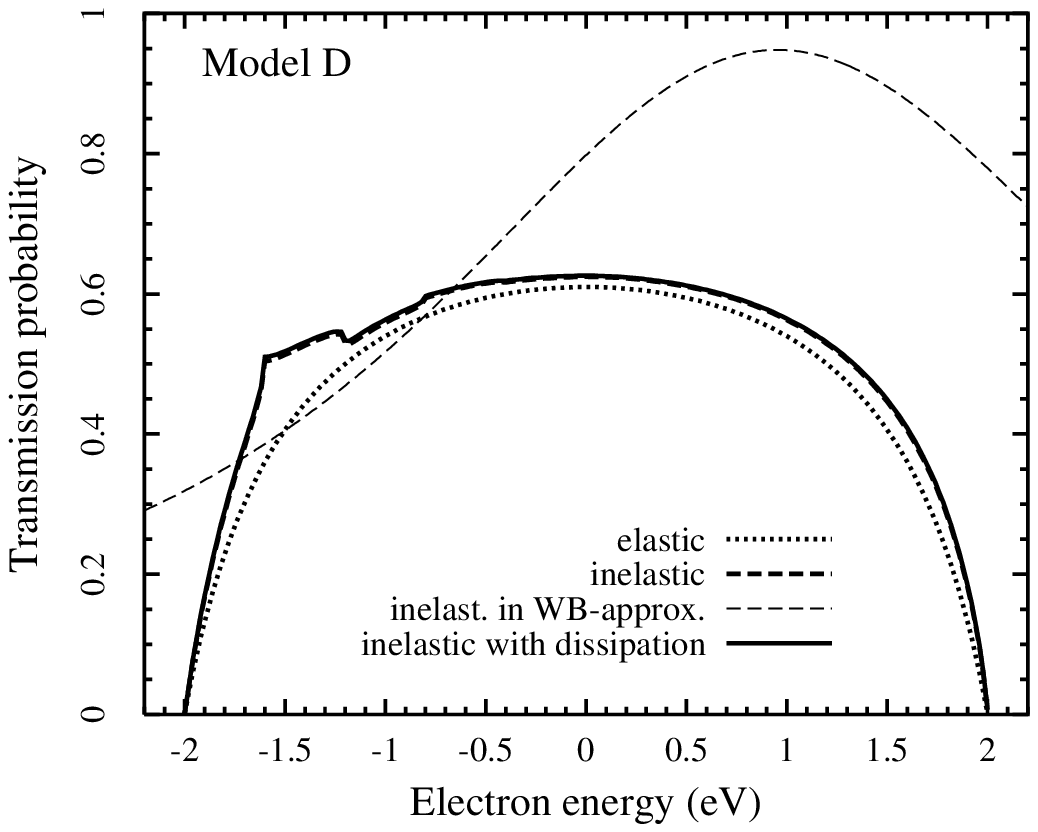}
 \end{center}
 \caption{\label{f:TallAa}
   Transmission probabilities for the case of
   a bridge strongly coupled to the leads (Model D) at zero bias voltage.}
\end{figure}

The results for the transmission probability are depicted in Fig.~\ref{f:TallAa}.
The strong coupling to the leads results in a rather broad transmission probability,
which is qualitatively well described taking into account
only elastic processes. Vibrationally inelastic contributions to the
 transmission probability manifest themselves
in various cusp structures. Thereby,  each cusp indicates the opening of 
a new vibrational channel. It is well-known from the theory of electron-molecule
scattering that the wide-band approximation is not at all applicable
in this case. 
The effect of the coupling to the bath, which is well described including the two 
lowest term in the expansion (\ref{tser}), is very small. This is a consequence
of the strong electronic coupling which results in a very short residence time
of the electron on the bridge. 

\begin{figure}
 \begin{center}
   \epsfxsize=0.49\textwidth \epsfbox{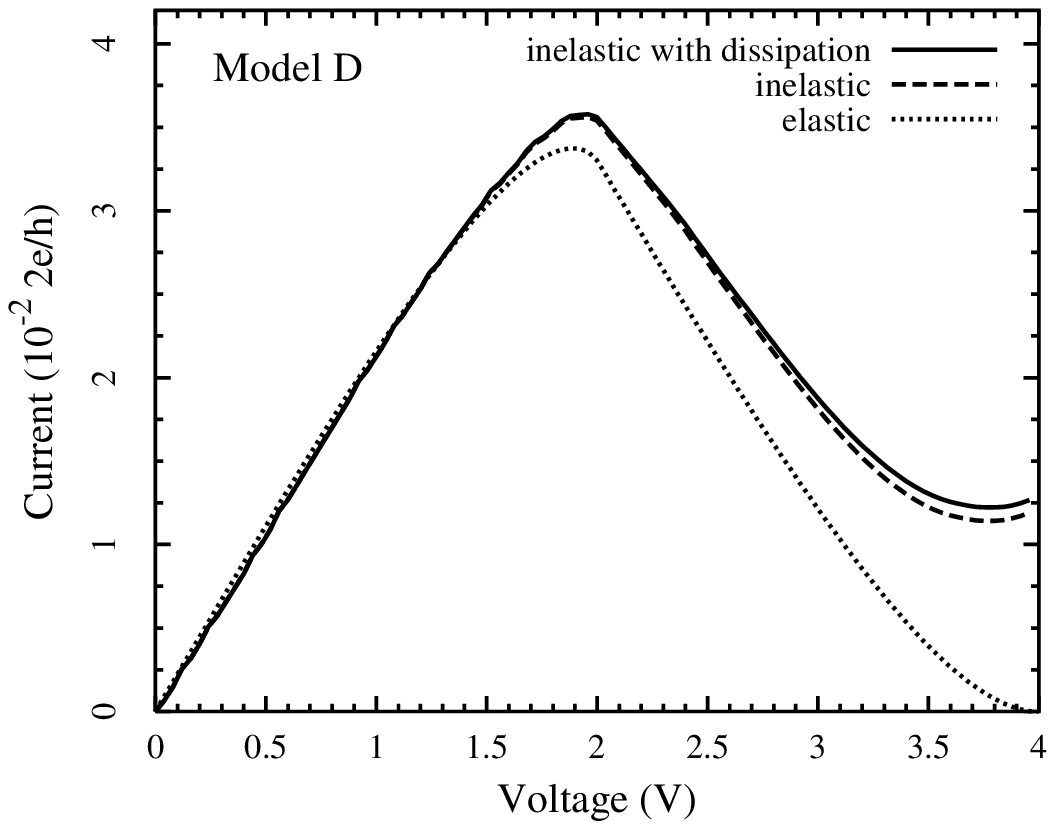}
 \end{center}
 \caption{\label{f:IVAa}
   Current-voltage characteristic for Model D.}
\end{figure}

Fig.~\ref{f:IVAa} displays the current-voltage characteristic for model D.
As to be expected from the transmission probabilities, the differences
among the various levels of theory are small with the exception 
of higher bias voltages. For higher voltages, the elastic current vanishes due 
to the empty overlap of the left and right conduction bands. 
Inelastic transmission processes, however, which are accompanied by
an energy loss of the electron, are still allowed.

\begin{figure}
 \begin{center}
   \epsfxsize=0.49\textwidth \epsfbox{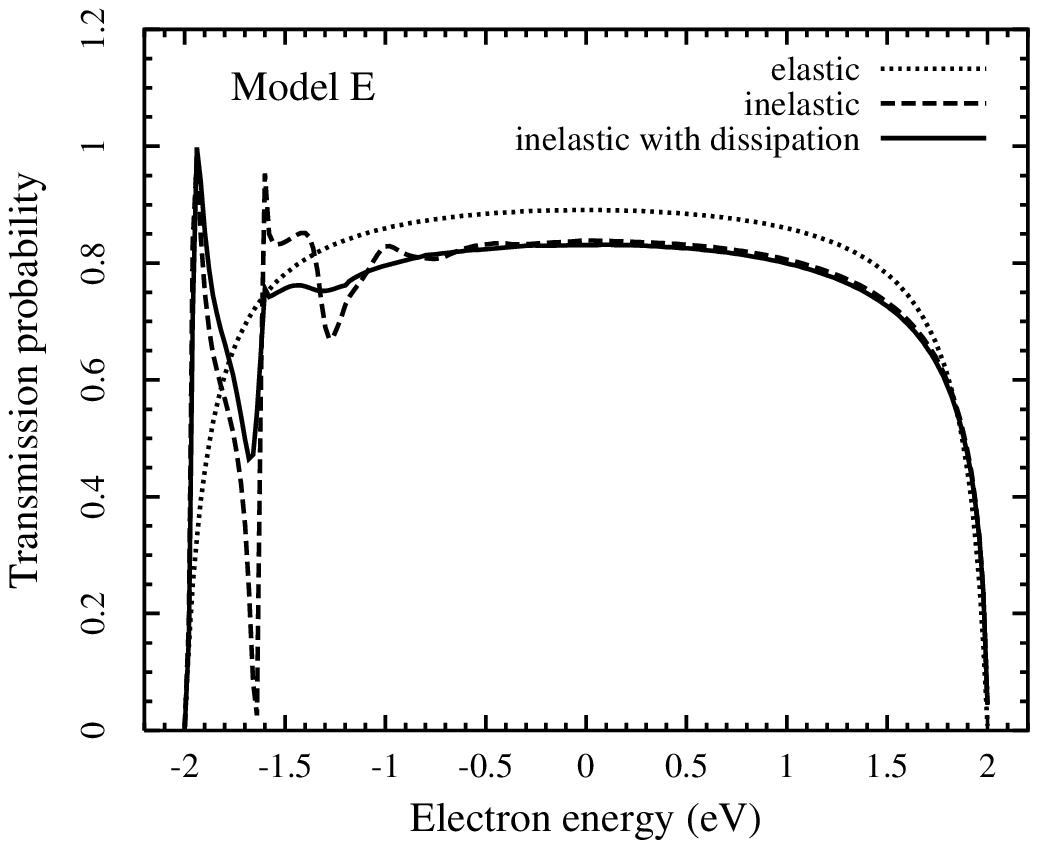}
 \end{center}
 \caption{\label{f:TallDa}
   Transmission probabilities for the case of
   a bridge strongly coupled to the leads (Model E) at zero bias voltage.}
\end{figure}

\begin{figure}
 \begin{center}
   \epsfxsize=0.49\textwidth \epsfbox{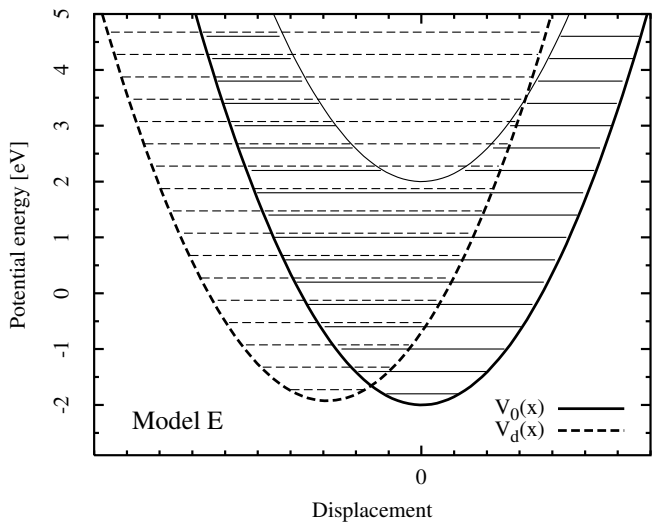}
 \end{center}
 \caption{\label{f:ModelD}
   Potential-energy curves for the interpretation 
   of the vibrational structures in Model E (as explained in the text).}
\end{figure}

\begin{figure}
 \begin{center}
   \epsfxsize=0.49\textwidth \epsfbox{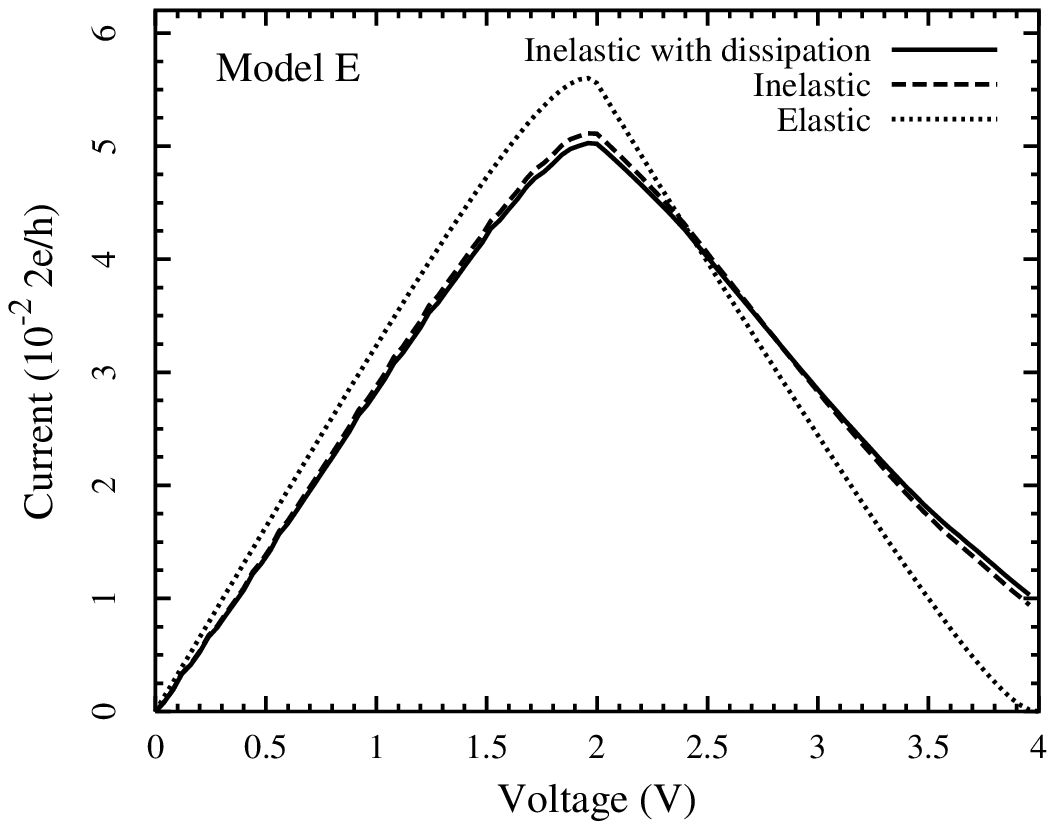}
 \end{center}
 \caption{\label{f:IVDa}
   Current-voltage characteristic for Model E}
\end{figure}

Although there are no structures in the transmission probability in 
Fig.\ \ref{f:TallAa} which are obviously related to the position of the discrete electronic state,
the location of this state does play an important role.
This is demonstrated in  Fig.~\ref{f:TallDa},
which shows the transmission probability  for Model E, which differs
from Model D only in a lower energy of the discrete electronic state, $\epsilon_d=-0.7$ eV. 
This different location of the discrete state results in pronounced peaks and 
minima in the transmission function, which are somewhat smoothed,
but not destroyed, by  the coupling to the bath. To facilitate the interpretation
of these structures, 
Fig.~\ref{f:ModelD} shows
the potential-energy curve of the discrete state 
(corresponding to $\tilde H_d$) together with 
the energies of the vibrational states. 
In addition, the potential energy of $\tilde H_0$ is shown,
shifted by $\pm 2$ eV, respectively, to indicate  the energy which
electrons coming from the conduction 
band may carry into the bridging molecule. Though the localized state is strongly coupled
to the continuum, due to the shift of the two potential 
curves, the ground vibrational state in $\tilde H_d$ has only a small 
overlap with the respective ground state in $\tilde H_0$. 
Consequently  the coupling between the two states is effectively small and sharp resonances 
may be observed. If the discrete state is localized higher in energy  
(as in model D), the potential 
energy curve of $\tilde H_d$ is shifted up. 
Then, the ground state of $\tilde H_d$ still has  
a small overlap with the ground state of $\tilde H_0$ but the decay into 
higher vibrational states becomes energetically possible. The sharp resonances thus 
"dissolve" in the continuum.
As is demonstrated in Fig.~\ref{f:IVDa}, the 
sharp structures close to the bottom of the conduction band have no
significant effect on the current, and the overall appearance 
of the current-voltage characteristic
is similar as in model D. 


\section{Conclusions}\label{sec:conclusion}

In this paper we have studied vibrationally inelastic effects on
electron transport through a molecular bridge that is connected to two metal
leads. The study was based on  a generic model for vibrational excitation
in resonant electron transmission processes through a molecular junction.
Employing projection-operator methods well-known
from resonant electron-molecule scattering, we have outlined how the transmission probability
can be evaluated numerically exactly within this model, 
without invoking the wide-band approximation or perturbation
theory with respect to the coupling between the bridging molecule and
the leads. Furthermore, the influence of dissipative vibrational processes
was investigated by considering the coupling of a vibrational reaction mode to
a dissipative bath.

The results of the model study can be summarized as follows:
In the case of tunneling through a molecular bridge which is weakly coupled to the leads,
the transfer of an electron may result in strong vibrational excitation,
which manifests itself in pronounced vibrational resonance structures in the transmission probability
and in a step-like appearance of the current-voltage characteristic.
Since in this case the residence time of the electron on the molecular bridge is relatively long,
dissipative processes such as vibrational relaxation can have a significant
effect on the dynamics. In particular, they result in a broadening of the resonance peaks
in the transmittance and of the step-like structures in the current-voltage characteristic. 
Furthermore, vibrational relaxation may
result in this case in a significant reduction of the overall magnitude of the current.
Due to the pronounced effects of the vibrational degrees of freedom, 
a theoretical treatment which only includes elastic
processes is not appropriate in this parameter regime. Our studies also show
that the wide-band approximation can only be applied if the
electronic resonance state is situated well within the conduction 
band and the electron-vibrational coupling is weak.

In the opposite case of a molecular bridge that is strongly coupled to the leads,
the transmission probability is typically characterized by a broad distribution, 
which in turn results in a rather structureless current-voltage characteristic. 
Nevertheless, the vibrational motion may manifest itself in cusp structures in the
transmittance. Furthermore, sharp resonance structures may occur in the transmission
probability, 
if the energy of the discrete electronic state
is low enough and the electronic-vibrational coupling sufficiently strong, 
such that the vibrational ground state of $\tilde H_d$ has some overlap with 
lower-lying vibrational states of $\tilde H_0$. 
Except for the latter case, the effect of vibrational relaxation is very small in this parameter regime.
The comparison of the results obtained at different levels of theory shows that
methods which only include elastic processes can give a rather good qualitative description 
of the electron transport in this case, although they miss the detailed cusp and resonance structures.
The wide-band approximation, on the other hand, 
is not valid in this parameter regime; due to the strong molecule-lead coupling, 
threshold effects become important which are neglected in the wide-band approximation.

To study the basic mechanisms of vibrationally inelastic electron transport,
we have focused in this work on relatively simple models with a  single 
harmonic reaction coordinate and a single electronic resonance state. 
It should be noted, however, that the  methods employed in this work are 
not limited to these models.
The extension of the theory to an anharmonic reaction coordinate, 
several reaction coordinates,  and several resonance states
is relatively straightforward.
Also, it should be emphasized that the potential-energy surfaces
of such models 
can in principle be determined {\em ab intio} by electronic structure
calculations. In this way, for example, the possibility of dissociation
of the molecular bridge induced by a strong current can be studied. 

Finally, it is noted that in the present work the current through
the molecular junction was obtained
with the generalized Landauer formula, Eq.\ (\ref{e:landauer}).
Although this formula  gives the correct description in the limit of weak coupling
between molecule and leads as well as in the situation when only single-electron processes
are important, it needs to be extended for applications where these assumptions are not fulfilled.
A theoretical treatment of inelastic processes without these limitations
is possible within the framework of non-equilibrium Green's function theory \cite{haug-jauho}.
The combination of this formalism with the methods employed in this 
paper is a challenging subject for future research.


\acknowledgments

Support of MC by a fellowship of the Alexander von Humboldt
Stiftung is gratefully acknowledged.


\begin{thebibliography}{10}

\bibitem{rzmbt97}
M.~A.\ Reed, C.\ Zhou, C.~J.\ Muller, T.~P.\ Burgin, and J.~M.\ Tour, 
Science {\bf 278},  252  (1997).

\bibitem{ruitenbeck02}
R.\ H.\ M.\ Smit, Y.\ Noat, C.\ Untiedt, N.\ D.\ Lang, M.\ C.\ van Hemert, 
and J.\ M.\ van Ruitenbeek, Nature {\bf 419}, 906 (2002).

\bibitem{reichert02}
J.\ Reicher, R.\ Ochs, H.B.\ Weber, M.\ Mayor, and H.\ von Lohneysen, 
Phys.\ Rev.\ Lett.\ {\bf 88}, 176804 (2002).

\bibitem{joachim00}
C.\ Joachim, J.\ K.\ Gimzewski, and A.\ Aviram, Nature {\bf 408}, 541 (2000).

\bibitem{n01}
A.\ Nitzan, Annu. Rev. Phys. Chem. {\bf 52},  681  (2001).

\bibitem{grossmann02}
F.\ Grossmann, R.\ Gutierrez, and R.\ Schmidt, Chem.\ Phys.\ Chem.\ {\bf 3}, 650 (2002).

\bibitem{nr03}
A.\ Nitzan and M.~A.\ Ratner, Science {\bf 300},  1384  (2003).

\bibitem{hry02}
P.\ H\"{a}nggi, M.\ Ratner, S.\ Yaliraki, Chem.\ Phys.\ {\bf 281}, 111 (2002).
{\em Special issue on: "Processes in molecular wires".}

\bibitem{dthrhk97}
S.\ Datta, W.\ Tian, S.\ Hong, R.\ Reifenberger, J.~I.\ Hederson, C.~P.\ Kubiak, 
Phys. Rev. Lett. {\bf 79},  2530  (1997).

\bibitem{ek98a}
E.~G. Emberly and G. Kirczenow, Phys. Rev. B {\bf 58},  10911  (1998).

\bibitem{tdhrhk98}
W.\ Tian, S.\ Datta, S.\ Hong, R.\ Reifenberger, J.~I.\ Henderson, C.~P.\ Kubiak, 
J. Chem. Phys. {\bf 109},  2874  (1998).

\bibitem{mj97}
M. Magoga and C. Joachim, Phys. Rev. B {\bf 56},  4722  (1997).

\bibitem{mj97b}
M. Magoga and C. Joachim, Phys. Rev. B {\bf 57},  1820  (1997).

\bibitem{mj97c}
M. Magoga and C. Joachim, Phys. Rev. B {\bf 59},  16011  (1997).

\bibitem{fclg97}
F. Faglioni, C.~L. Claypool, N.~S. Lewis, and W.~A. Goddard, J. Phys. Chem.
  {\bf 101},  5996  (1997).

\bibitem{lk99}
S. Larsson and A. Klimkans, Theochem.\ J.\ Mol.\ Struc. {\bf 464},  59  (1999).

\bibitem{yrgmr99}
S.~N.\ Yaliraki, A.~E.\ Roitberg, C.\ Gonzalez, V.\ Mujica, M.~A.\ Ratner,
J. Chem. Phys. {\bf 111},  6997  (1999).

\bibitem{l95}
N.~D. Lang, Phys. Rev. B {\bf 52},  5335  (1995).

\bibitem{ht95}
K. Hirose and M. Tsukada, Phys. Rev. B {\bf 51},  5278  (1995).

\bibitem{la00}
N.~D. Lang and P. Avouris, Phys. Rev. Lett. {\bf 84},  358  (2000).

\bibitem{vpl00}
M. {Di Ventra}, S.~T. Pantelides, and N.~D. Lang, Phys. Rev. Lett. {\bf 84},
  979  (2000).

\bibitem{mrr00}
V. Mujica, A.~E. Roitberg, and M.~A. Ratner, J. Chem. Phys. {\bf 112},  6834
  (2000).

\bibitem{lb03}
B.\ Larade, and A.~M.\ Bratkovsky, Phys.\ Rev.\ B {\bf 68}, 235305 (2003). 

\bibitem{landauer57}
R.\ Landauer, IBM J.\ Res.\ Dev.\ {\bf 1}, 223 (1957).

\bibitem{landauer81}
R.\ Landauer, Phys.\ Lett.\  A {\bf 85}, 91 (1981).

\bibitem{datta}
S. Datta, {\em Electric Transport in Mesoscopis Systems} (Cambridge University
  Press, 1995).

\bibitem{pplaam00}
H.\ Park, J.\ Park, A.~K.~L.\ Lim, E.~H.\ Anderson, A.~P.\ Alivisatos, P.~L.\ McEuen, 
Nature {\bf 407},  57  (2000).

\bibitem{omwkrlm98}
M.\ Olson, Y.\ Mao, T.\ Windus, M.\ Kemp, M.\ Ratner, N.\ L\'{e}on, V.\ Mujica, 
J. Phys. Chem. {\bf B 102},  941  (1998).

\bibitem{s03}
W. Schmickler, Chem. Phys. {\bf 289},  349  (2003).

\bibitem{trn03}
A. Troisi, M.~A. Ratner, and A. Nitzan, J. Chem. Phys. {\bf 118},  6072
  (2003).

\bibitem{ek00}
E.~G. Emberly and G. Kirczenow, Phys. Rev. B {\bf 61},  5740  (2000).

\bibitem{w03}
K. Walczak, arXiv: cond-mat/0306174  1  (2003).

\bibitem{bs01}
D. Boese and H. Schoeller, Europhys. Lett. {\bf 54},  668  (2001).

\bibitem{mpt03}
K.~D. {McCarthy}, N. {Prokof'ev}, and M.~T. Tuominen, Phys. Rev. B {\bf 67},
  245415  (2003).

\bibitem{bf04}
S.\ Braig, and K.\ Flensberg, Phis. Rev. B {\bf 68}, 205324 (2004).

\bibitem{givksj98}
L.~Y.\ Gorelik, A.\ Isacsson, M.~V.\ Voinova, B.\ Kasemo, R.~I.\ Shekhter, and M.\ Jonson,
Phys. Rev. Lett. {\bf 80}, 4526 (1998).

\bibitem{ndj03}
T.\ Novotn\'{y}, A.\ Donarini, A.-P.\ Jauho, Phys. Rev. Lett. {\bf 90}, 256801 (2003).

\bibitem{segal02}
D.\ Segal and A.\ Nitzan, J.\ Chem.\ Phys.\ {\bf 117}, 3915 (2002).

\bibitem{segal03}
D.\ Segal, A.\ Nitzan, and P.\ H\"anggi, J.\ Chem.\ Phys.\ {\bf 119}, 6840 (2003).

\bibitem{yssb99}
Z.~G. Yu, D.~L. Smith, A. Saxena, and A.~R. Bishop, Phys. Rev. B {\bf 59},
  16001  (1999).

\bibitem{nf99}
H. Ness and A.~J. Fisher, Phys. Rev. Lett. {\bf 83},  452  (1999).

\bibitem{d84}
F.~I. Dalidchik, Sov. Phys. JETP {\bf 60},  795  (1984).

\bibitem{shls91}
J.~A. Stovneng, E.~H. Hauge, P. {Lipavsk\'{y}}, and V. \v{S}pi\v{c}ka, Phys.
  Rev. B {\bf 44},  13595  (1991).

\bibitem{hb99}
K. Haule and J. {Bon\v{c}a}, Phys. Rev. B {\bf 59},  13087  (1999).

\bibitem{wjw88}
N.~S. Wingreen, K.~W. Jacobsen, and J.~W. Wilkins, Phys. Rev. Lett. {\bf 61},
  1396  (1988).

\bibitem{wjw89}
N.~S. Wingreen, K.~W. Jacobsen, and J.~W. Wilkins, Phys. Rev. B {\bf 40},
  11834  (1989).

\bibitem{pb87}
B.~N.~J. Persson and A. Baratoff, Phys. Rev. Lett. {\bf 59},  339  (1987).

\bibitem{lp00}
N. Lorente and M. Persson, Phys. Rev. Lett. {\bf 85},  2997  (2000).

\bibitem{sb97}
C. Spataru and P. Budau, J. Phys.: Condensed Matter {\bf 9},  8333  (1997).

\bibitem{ga93}
M.~A. Gata and P.~R. Antoniewicz, Phys. Rev. B {\bf 47},  13797  (1993).

\bibitem{bhg93}
A. Bringer, J. Harris, and J.~W. Gadzuk, J. Phys.: Condensed Matter {\bf 5},
  5141  (1993).

\bibitem{bh94}
M. Brandbyge and P. {Hedeg{\aa}rd}, Phys. Rev. Lett. {\bf 72},  2919  (1994).

\bibitem{gpl97}
S. Gao, M. Persson, and B.~I. Lundqvist, Phys. Rev. B {\bf 55},  4825  (1997).

\bibitem{ga99}
M.~A. Gata and P.~R. Antoniewicz, Phys. Rev. B {\bf 60},  8999  (1999).

\bibitem{clkh03}
S.\ Camalet, J.\ Lehmann, S.\ Kohler, and P.\ H\"{a}nggi, 
Phys. Rev. Lett. {\bf 90}, 210602 (2003).

\bibitem{lkhn03}
J.\ Lehmann, S.\ Kohler, P.\ H\"{a}nggi, and A.\ Nitzan,
J.\ Chem.\ Phys.\ {\bf 118}, 3283 (2003).

\bibitem{s73b}
G.~J. Schulz, Rev. Mod. Phys. {\bf 45},  423  (1973).

\bibitem{a89}
M. Allan, J.\ El.\ Spectrosc.\ {\bf 48},  219
   (1989).

\bibitem{d91}
W. Domcke, Phys. Rep. {\bf 208},  97  (1991).

\bibitem{chad02}
M. \v{C}\'{\i}\v{z}ek, J. Hor\'{a}\v{c}ek, M. Allan, and W. Domcke, Czech.\ J.
  Phys. {\bf 52},  1057  (2002).

\bibitem{gsl89}
B.~Y. Gelfand, S. Schmitt-Rink, and A.~F.~J. Levi, Phys. Rev. Lett. {\bf 62},
  1683  (1989).
\bibitem{Feshbach62}
H.~Feshbach, Ann.\ Phys.\ {\bf 19}, 287 (1962).

\bibitem{note1}
See, for example Ref.\ \cite{n01}, page~724.

\bibitem{td98}
M. Thoss and W. Domcke, J. Chem. Phys. {\bf 109},  6577  (1998).

\bibitem{kbpevmj99}
C.~Kerqueris, J.-P.\ Bourgoin, S.~Palacin, D.~Esteve, C.~Urbina, M.~Magoga, C.~Joachim, 
Phys. Rev. B {\bf 59},  12505  (1999).

\bibitem{bmd85}
M. Berman, C. {M\" undel}, and W. Domcke, Phys. Rev. A {\bf 31},  641  (1985).

\bibitem{economou}
E.~N. Economou, {\em Green's Functions in Quantum Physics} (Springer-Verlag, 1983).

\bibitem{ds}
D.~S. M~C~{Desjonqu\`{e}res}, {\em Concepts in Surface Physics}
  (Springer-Verlag, 1993).

\bibitem{Hopkins80}
J.\ B.\ Hopkins, D.\ E.\ Powers, and R.\ E.\ Smalley, J.\ Chem.\
Phys.\ {\bf 72}, 5093 (1980).

\bibitem{Mukamel80}
S.\ Mukamel and R.\ E.\ Smalley, J.\ Chem.\ Phys.\ {\bf 73}, 4156
(1980). 

\bibitem{Freed83}
K.\ F.\ Freed and A.\ Nitzan, in: {\it Energy Storage and
Redistribution in Molecules}, ed. J.\ Hinze (Plenum, New York, 1983),
p.\ 467.

\bibitem{Brumer91}
D.\ Gruner and P.\ Brumer, J.\ Chem.\ Phys.\ {\bf 94}, 2848 (1991).

\bibitem{Uzer91}
T.\ Uzer, Phys.\ Rep.\ {\bf 199}, 124 (1991).

\bibitem{Zewail94}
A.\ H.\ Zewail, {\it Femtochemistry--Ultrafast Dynamics of the
Chemical Bond}, (World Scientific, Singapore 1994), Vol.\ II, p.\ 703.

\bibitem{Nesbitt96}
D.\ J.\ Nesbitt and R.\ W.\ Field, J.\ Phys.\ Chem.\ {\bf 100}, 12735
(1996).

\bibitem{Leggett87}
A.~J.~Leggett, S.~Chakravarty, A.~T.~Dorsey, M.~P.~A.~Fisher, A.~Garg,
and W.~Zwerger, Rev.\ Mod.\ Phys.\ {\bf 59}, 1 (1987).

\bibitem{Weiss99}
U.\ Weiss, {\it Quantum Dissipative Systems} (World Scientific,
Singapore, 1999).

\bibitem{w00b}
M.\ Wagner, Phys. Rev. Lett. {\bf 85},  174 (2000).

\bibitem{s92}
F.\ Sols, Ann. Phys. {\bf 214}, 386 (1992).

\bibitem{Cizekunp}
M.\ \v{C}\'{\i}\v{z}ek, unpublished.

\bibitem{f04} 
K.\ Flensberg, Phys. Rev. B {\bf 68}, 205323 (2004).

\bibitem{haug-jauho}
H. Haug and A.~P. Jauho, {\em Quantum Kinetics in Transport and Optics of
  Semi-conductors} (Springer-Verlag, 1998).

\bibitem{nf02}
H.\ Ness, A.~J.\ Fisher, Europhys.\ Lett.\ {\bf 57}, 885 (2002).

\bibitem{h95}
J. {Hor\'{a}\v{c}ek}, J. Phys. B {\bf 28},  1585  (1995).

\bibitem{ab03}
A.~S. Alexandrov, A.~M. Bratkovsky, Phys. Rev. B {\bf 67}, 235312 (2003).

\bibitem{chm00}
M. {\v C\'\i\v zek}, J. {Hor\' a\v cek}, and H.-D. Meyer, Comput. Phys.
  Communications {\bf 131},  41  (2000).

\bibitem{dc80}
W. Domcke and L.~S. Cederbaum, J. Phys. B {\bf 14},  149  (1980).

\bibitem{mhc91}
{H.-D. Meyer}, J. {Hor\'{a}\v{c}ek}, and L.~S. Cederbaum, Phys. Rev. A {\bf
  43},  3587  (1991).

\bibitem{bp88}
A. Baratoff and B.~N.~J. Persson, J. Vac. Sci. Technol. A {\bf 6},  331
  (1988).

\bibitem{gs88}
L.~I. Glazman and R.~I. Shekhter, Sov. Phys. JETP {\bf 67},  163  (1988).

\end{thebibliography}


\begin{table}
 \begin{center}
  \begin{tabular}{ccccc}
    Model & $\epsilon_{d}$ & $v$ & $\omega_{\rm S}$ & $\lambda$ \\
  \hline
      A   &      0.5       & 0.2 &       0.5        &    0.3    \\
      B   &     -0.5       & 0.2 &       0.5        &    0.3    \\
      C   &      0.5       & 0.2 &       0.4        &    0.7    \\
  \hline
      D   &      1.6       &  1  &       0.4        &    0.7    \\
      E   &     -0.7       &  1  &       0.4        &    0.7  
  \end{tabular}
 \end{center}
\caption{\label{t:modpar}
  Parameters for the different models considered. We have set
   $\beta=1$ eV in all cases, and $\eta=0.1$, 
  $\omega_{\rm C}=\omega_{\rm S}$ if the coupling to the dissipative bath is taken into 
  account. Furthermore, the center of the conduction band 
  is equal to Fermi energy $\mu_{\rm L/R}=\pm{\frac{1}{2}}V=\epsilon_{\rm F}$ 
  for the left and right lead respectively}
\end{table}


\end{document}